\documentclass[a4paper,fleqn]{cas-dc}

\usepackage{graphicx}

\usepackage{subcaption}

\usepackage{algorithm}
\usepackage{algpseudocode}
\usepackage{cleveref}

\usepackage{soul}
\usepackage{tikz}
\usepackage{ifthen}

\usepackage{pgfplots}
\pgfplotsset{
    compat=1.18, 
    mystyle/.style={ 
        axis lines=middle,
        grid=major,
        xlabel={Eje X},
        ylabel={Eje Y},
        every axis x label/.append style={below}, 
        every axis y label/.append style={left},  
        legend pos=north west, 
    },
    line style/.style={ 
        thick,
        smooth,
        red,
    },
}

\usepackage[authoryear,longnamesfirst]{natbib}

\algblockdefx[kernel]{BKernel}{EKernel}{\textbf{For Parallel}}

\def\tsc#1{\csdef{#1}{\textsc{\lowercase{#1}}\xspace}}
\tsc{WGM}
\tsc{QE}

\usepackage{xcolor}

\begin{document}
\let\WriteBookmarks\relax
\def\floatpagepagefraction{1}
\def\textpagefraction{.001}

\shorttitle{RT Filter 3D Convex Hulls}    

\shortauthors{Carrasco et~al.}  

\title [mode = title]{Convex Hull 3D Filtering with GPU Ray Tracing and Tensor Cores}  



%

\author[1]{Roberto Carrasco}[]



\ead{rocarras@dcc.uchile.cl}



\affiliation[1]{organization={Departamento Ciencias de la Computaci\'on, Universidad de Chile.},
    city={Santiago},
    country={Chile}}

\author[2]{Enzo Meneses}[]

\author[2]{H\'ector Ferrada}[]

\author[2]{Crist\'obal A. Navarro}[]


\affiliation[2]{organization={Instituto de Inform\'atica Universidad Austral de Chile},
    city={Valdivia},
    country={Chile}}

\author[1]{Nancy Hitschfeld}


\begin{abstract}
In recent years, applications such as real-time simulations, autonomous systems, and video games increasingly demand the processing of complex geometric models under stringent time constraints. Traditional geometric algorithms, including the convex hull, are subject to these challenges. A common approach to improve performance is scaling computational resources, which often results in higher energy consumption. Given the growing global concern regarding sustainable use of energy, this becomes a critical limitation.
This work presents a 3D preprocessing filter for the convex hull algorithm using ray tracing  and tensor core technologies. The filter builds a delimiter polyhedron based on Manhattan distances that discards points from the original set. The filter is evaluated on two point distributions: uniform and sphere. Experimental results show that the proposed filter, combined with convex hull construction, accelerates the computation of the  3D convex hull by up to $200 \times$ with respect to a CPU parallel implementation.
This research demonstrates that geometric algorithms can be accelerated through massive parallelism while maintaining efficient energy utilization. Beyond execution time and speedup evaluation, we also analyze GPU energy consumption, showing that the proposed preprocessing filter not only reduces the computational workload but also achieves performance gains with controlled energy usage. These results highlight the dual benefit of the method in terms of both speed and energy efficiency, reinforcing its applicability in modern high-performance scenarios.
\end{abstract}




\maketitle

\section{Introduction}

Currently, many applications have emerged that require processing complex geometric models in very limited time or in real time, such as simulations or video games, where the traditional geometry algorithm problems, such as the convex hull algorithm, are not alien to this problem. Often, the solution to speed up an algorithm is to increase processing resources, but this incurs an increase in energy consumption. In turn, we have observed that the world has become much more aware of how energy resources are used. This research demonstrates that it is possible to accelerate an algorithm through massive parallelism with efficient use of available energy resources.

The Graphics Processing Unit (GPU) has become an essential asset for accelerating applications across several fields, such as science, technology, and entertainment, due to its highly efficient parallel computing architecture. GPUs leverage a general-purpose programming model designed for developing scalable parallel algorithms, making it possible to accelerate solutions for data-parallel problems. In Nvidia's CUDA platform, four key components structure this model: the kernel, which is the parallel code designed by the programmer to run on the GPU, and the hierarchical organization of resources into threads, blocks, and grids. Each thread executes the kernel, with threads grouped into blocks and blocks organized into a grid, defining the complete resource structure to execute the parallel code.

Recent advancements in GPU architecture have introduced specialized cores, such as tensor cores and ray-tracing cores, which significantly enhance performance for specific tasks. Tensor cores (TC) are designed to accelerate AI-based computations, particularly matrix operations critical for deep learning. Ray-tracing (RT) cores, on the other hand, are specialized for real-time ray-tracing, enabling more realistic lighting, shadows, and reflections in graphics by simulating the physical behavior of light. Integrating these cores into modern GPUs has expanded their capability beyond traditional parallel tasks, offering dedicated acceleration for advanced workloads in both scientific and graphical applications.

On the other hand, there are fundamental problems in computer science and computational geometry that can take advantage of modern GPUs, especially when  large data set are processed (in real time or not). One of these problems is the computation of convex hull in high dimensions. The convex hull   is a fundamental geometric concept with many applications in computer graphics, robotics, data mining, and other fields \cite{MEERAN1997737, Nearchou_1994, NEMIRKO2021381}. This work accelerates the computation of the 3D convex hull which is defined as the algorithm that computes the smallest convex polyhedron that encloses a set of points in three-dimensional space \cite{Berg:2008:CGA:1370949}. 

 Although the RT cores were originally designed for achieving hardware-acceleration of graphical tasks, recent research has found ways to leverage them on applications outside graphics \cite{zhu2022rtnn, meneses2024accelerating, zhao2023leveraging}, with very positive performance results. In this context, reformulating the convex hull filtering process as a ray tracing problem could potentially lead to significative performance gains. Based on a previous research that proposed a traditional GPU filter for the 2D convex hull \cite{CARRASCO2024104793}, this work goes one step further and proposes a 
 filter for the 3D convex hull using both RT and Tensor cores.

The remainder of this manuscript is organized as follows: Section \ref{sec:related_work} reviews existing literature and alternative approaches for accelerating convex hull computation, while Section \ref{sec:overview} provides a technical overview of the CUDA, Tensor, and Ray Tracing cores utilized in this study. 
Section \ref{sec:algorithm} details the expansion into 3D preprocessing filters \cite{CARRASCO2024104793}, leveraging modern GPU hardware. Section \ref{sec:experiment} presents the experimental evaluation, where the algorithm’s efficiency is assessed across the uniform, and sphere point distributions. Finally, Section \ref{sec:conclus} offers concluding remarks and summarizes the preliminary findings, which demonstrate that the proposed filter achieves speedups of up to 200$\times$ over existing implementations.

\section{Related Work}
\label{sec:related_work}
Convex hull algorithms have been studied for many years and many different algorithms and techniques have been developed to improve their efficiency and accuracy. However, there is still a need for faster and more effective convex hull algorithms, especially for solving large-scale problems in real time \cite{o1998computational}.
The most popular algorithms for computing the convex hull of a set of points  in 2D are  the quickhull algorithm \cite{preparata1993computational}, with  worst case time complexity of $O(n^2)$ and an average time complexity of $O(n \log n)$,   the incremental algorithm \cite{KALLAY1984197} with a worst-case time complexity of $O(n^2)$, the divide-and-conquer algorithm \cite{10.1145/359423.359430} with a time complexity of $O(n \log n)$ and the gift wrapping algorithm with a time complexity of $O(nh)$ \cite{10.1145/321556.321564,JARVIS197318}, where $h$ is the number of point in the convex hull.

Current approaches are on the optimization of data sets used as input to traditional convex hull computation algorithms, through preprocessing filters that discards points that are not candidates to be part of the hull using basic operations with low computational cost. In this section, we divide the state-of-the-art into  sequential CPUs parallel CPU, and GPU algorithms, where  optimization strategies to improve the performance and efficiency for 2D and 3D problems are described.

\subsection{Sequential algorithms}
Many convex hull algorithms have a strong dependency on the size of the point set input; a solution to this is preprocessing algorithms that can reduce the input size by $O(n)$, discarding points that are not part of the hull with fast operations \cite{ CUDAch2d,CUDAch3d} before calculating the convex hull. The most widely used and efficient method to improve computational performance is to eliminate interior points that are not candidates for the hull.  

The quick-hull algorithm \cite{preparata1993computational} was the first one  that introduced the idea of using a filter (quadrilateral  defined by four extreme points) to discard in O(n), the points that are not candidates to the  hull. More recently, the  work developed by Skala et al. \cite{skala01,skala02,skala03} built a approximate quadrilateral from a sample of 10\% of the input set points, and this quadrilateral is used to  discard the points that are no candidates to the hull. The algorithm  works subdividing the space using polar coordinates. They used a simplifier calculation of the approximated angle to accelerate the quadrant search.
Alshamrani et al. \cite{alshamrani} extend the idea of quick-hull and  built a polygon formed by 8 extreme  points and thy used a priority queue to keep the candidate points. They report an algorithm up to $77 \times$ and $12 \times$ faster than the Graham scan and Jarvis algorithms, respectively. This work currently does not have a 3D or GPU implementation reported at the time of writing this work.  Ferrada et al. \cite{FERRADA2020112298} developed \texttt{heaphull} a fast and compacted (memory-efficient) implementation for discarding points in 2D convex hulls using the Manhattan distance as the primary metric. They reported a speedup of $1.7 \times$ to $10 \times$ faster than convex hull methods available in the CGAL library.

Some applications require the construction of adaptive convex hulls.  When new points appear, a new hull is computed by merging previous hulls in the plane \cite{10.1007/978-3-319-94776-1_14}. This type of strategy is highly parallelizable to accelerate convex hull algorithms because multiple hulls can be computed in parallel and then interlace whole hulls, however, this work does not currently have a parallel implementation.

The research to improve sequential  algorithms for convex hull computation continues to evolve, offering increasingly efficient and accurate solutions to a wide range of problems. Two of the most efficient sequential implementations of the convex hull are provided by the Qhull \cite{10.1145/235815.235821} and CGAL \cite{cgal:hs-ch3-18b} libraries, which implement one or more of the algorithms mentioned above.


\subsection{Parallel CPU strategies}

ParGeo is a multi-core library for computational geometry, developed by Wang \cite{ParGeo}, that provides several modules for KD-tree-based spatial search, spatial graph generation, and fundamental computational geometry algorithms. One of these fundamental algorithms is the computation of the convex hull. The library provides five implementations: brute force, serial quick hull, parallel divide and conquer, parallel quick-hull, and parallel Pseudohull. The Pseudohull implementation is based on a reservation technique that facilitates simultaneous modifications, ensuring the hull can be processed in parallel; this is  the fastest convex hull implementation: it achieves up to $43.7 \times$ parallel speedup compared to Qhull algorithm\cite{10.1145/235815.235821}, and up to $5.05 \times$ over CGAL implementations on 3-dimensional shapes.

\subsection{Parallel GPU strategies}

Another way to accelerate convex hull algorithms is by using massive parallelism such as that provided by GPU technology.
Srungarapu et al. \cite{5763404} accelerated  a 2D quick-hull based algorithm by parallelizing in the GPU the identification of extreme points, the tagging of points residing within the polygon, and the scanning operations; however, the main loop is in CPU. Their research findings indicate a  speedup of up to $14 \times$ compared to the Qhull to solve convex hull problems, as documented in their work .

Blelloch et al. \cite{10.1145/3350755.3400255} present a theoretical analysis on a parallel incremental randomized algorithm.  This work is an $O(\log n)$ dependence depth with high probability. However, this work only exists in 2D data and an implementation is not available yet, as of the time of writing this manuscript, no practical implementation has been offered. 

Mei developed rotational filtering techniques to compute the convex hull in 2D and 3D \cite{mei2012,mei} . They rotate the entire plane to obtain the extreme points at each rotation. Subsequently, they use a preprocessing approach to classify all the points and discard those that do not belong to the convex hull on the GPU. This preprocessing resulted in up to a $6 \times$ speedup compared to  Qhull. 

In their work, Stein et al. developed a parallel algorithm to calculate the 3D convex hull of a set of points, using the CUDA programming model \cite{STEIN2012265}. Their method, built on the quickhull technique in GPU parallel, demonstrated a remarkable performance boost, achieving a speedup of $30 \times$ compared to  Qhull. As of today, there is no implementation of this approach available for download.

Keith et al. \cite{alanhull}  developed a GPU implementation inspired in the \texttt{heaphull} algorithm \cite{FERRADA2020112298}.  The authors reported   $4.4 \times$ faster than the sequential  \texttt{heaphull} algorithm and $3.2 \times$ faster than other existing GPU-based approaches in the state-of-the-art. However, Ferrada's algorithm \cite{FERRADA2020112298} only counts with a 2D implementation.

Sequential 2D algorithms that includes a filtering step have the potential to be accelerated using the GPU programming model and, at the same time, 2D parallel strategies can be adapted/extended to support more dimensions. This presents an opportunity to enhance current filtering methods by incorporating the GPU programming model for solving 3D problems. 

\section{Overview of Modern GPUs}
\label{sec:overview}
Modern GPUs have been incorporating new features over time, starting with CUDA cores around 2007, and continuing with the two most relevant ones for this work, which are the Tensor cores and Ray Tracing cores, in recent years.

\subsection{CUDA Cores}
The NVIDIA CUDA cores are the default cores that are used by any regular CUDA program for GPU. Released in 2007, these cores correspond to the classic INT32/FP32 units with their basic arithmetic operations such as addition, product, and also some special elementary functions (although more limited in quantity). The CUDA API is an extension of the C language for coding general-purpose programs, providing a programming model that we currently use to program on GPUs. This programming model plays an important role in the design and development of GPU-accelerated programs and is governed by a three-level hierarchy that defines the GPU's parallel workspace. This hierarchy corresponds to thread, block, and grid which provide the threads for the compute kernel.

\subsection{Tensor Cores}
Due to the fast adoption of deep learning in multiple fields of science and technology, CPU and GPU manufacturing companies have started including application-specific integrated circuits (ASICs) to their processors to further accelerate the computational tasks involved in the phases of training and inference in DLSS (Deep Learning Super Sampling) applications. This change has led to the inclusion of Tensor Core (TC) units in recent Nvidia GPUs, which are specific-purpose processing units that sit next to the GPU CUDA cores in the streaming multiprocessors (SM) of the chip. With tensor cores, the operation exposed to the programmer is the matrix-multiply-accumulate (MMA), defined as
\begin{equation}
    D_{m \times n} = A_{m \times k} \times B_{k \times n} + C_{m \times n}
\end{equation}
the programming model exposes the MMA operation in terms of dimensions $m \times n \times k$, where each matrix has a specific size depending on the numerical precision. For instance, with FP16 precision the matrices have $16 \times 16 = 256$ elements. An algorithm that takes advantage of the acceleration of the tensor core would redesign its work to include multiple MMA operations that occur in $O(1)$ accelerated by hardware in parallel.

\subsection{Ray Tracing Cores}
Ray tracing (RT) cores were developed primarily for real-time computer graphics, particularly for applications in ray tracing, although its versatility extends beyond that. The RT pipeline was crafted to offer a streamlined abstraction layer for developing hardware-accelerated lighting algorithms, with ray tracing (and path tracing) being a key focus. Ray tracing simulates the behavior of light as it interacts with objects and surfaces within a scene, creating highly realistic images that closely mimic the physical world. Due to the significant computational demands of ray tracing, many advancements have focused on optimizing its efficiency, with the Bounding Volume Hierarchy (BVH) serving as an effective acceleration structure (AS) that is implemented in hardware in the RT core.

The BVH organizes the geometry of the scene in a hierarchical tree structure that contains multiple boundaries, efficiently guiding rays through the scene by reducing collision checks. In the ray tracing process, rays are projected from a given point into the scene and traverse the BVH until they intersect a geometric primitive, such as a sphere, square, or triangle. Historically, ray tracing was computationally prohibitive for real-time applications, and as a result, raster-based methods were preferred for real-time rendering. Today, ray tracing operations are $O(\log n)$ for BVH traversal and $O(1)$ for ray intersections, both hardware-accelerated. However, in 2018, Nvidia introduced RT cores within its Turing architecture, specifically designed to accelerate BVH traversal and ray/triangle intersection testing. This innovation enabled real-time ray tracing, with RT cores sitting alongside CUDA and Tensor cores on modern GPUs. 

To program with ray tracing, there exist APIs such as NVIDIA OptiX, Microsoft DXR, and Vulkan, which are optimized for RT core acceleration. OptiX, the API used in this work, organizes ray tracing into an eight-stage programmable pipeline. The Ray Generation shader, executed at the start of the pipeline, generates rays from a specified 3D point in space. A payload, corresponding to the information that you want to travel with the rays, defined by the user, can be attached to the beam to access it in several stages. The Intersections stage is optimized by OptiX and RT cores to manage BVH traverse and triangle intersections. The Miss and Closest-hit shaders, which are programmable, activate when a ray encounters no collisions or when the closest collision is detected, respectively. The Any-hit shader, also programmable, activates upon each collision and can be bypassed to improve performance. The Direct Callable and  Continuation Callable stages provide custom routines for finalizing calculations or launching new rays.

\section{Algorithm Design and Implementation}
\label{sec:algorithm}
\begin{figure*}[htb] 
  \centering
  \begin{subfigure}{0.24\textwidth}
    \centering
    \includegraphics[width=\linewidth]{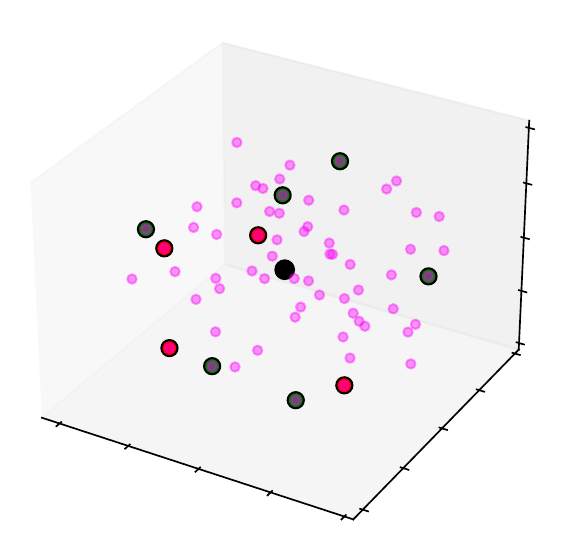}
    \caption{Phases 1 and 2}
    \label{fig:input}
  \end{subfigure}
   \hfill
  \begin{subfigure}{0.24\textwidth}
    \centering
    \includegraphics[width=\linewidth]{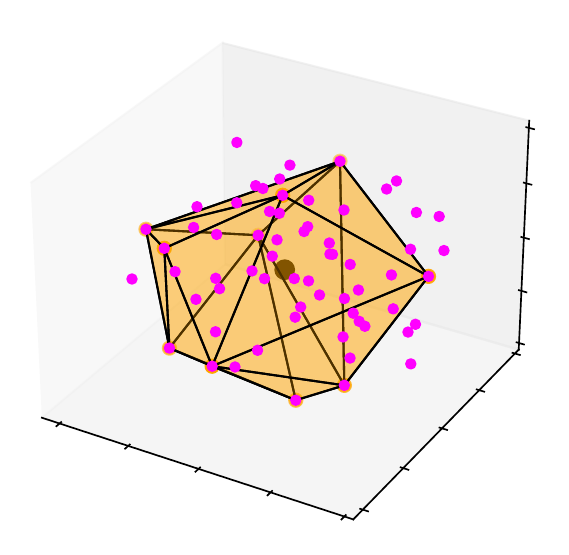}
    \caption{Phase 3}
    \label{fig:poly_filter}
  \end{subfigure}
   \hfill
  \begin{subfigure}{0.24\textwidth}
    \centering
     \includegraphics[width=\linewidth]{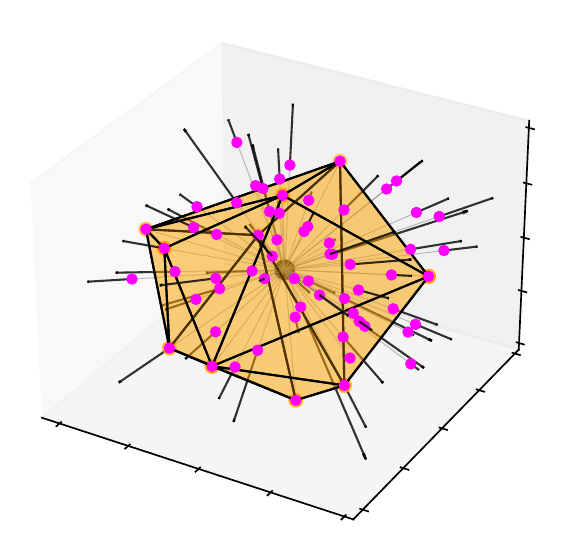}
    \caption{Phases 4 and 5}
    \label{fig:convexhull_sphere}  
  \end{subfigure}
   \hfill
  \begin{subfigure}{0.24\textwidth}
    \centering
    \includegraphics[width=\linewidth]{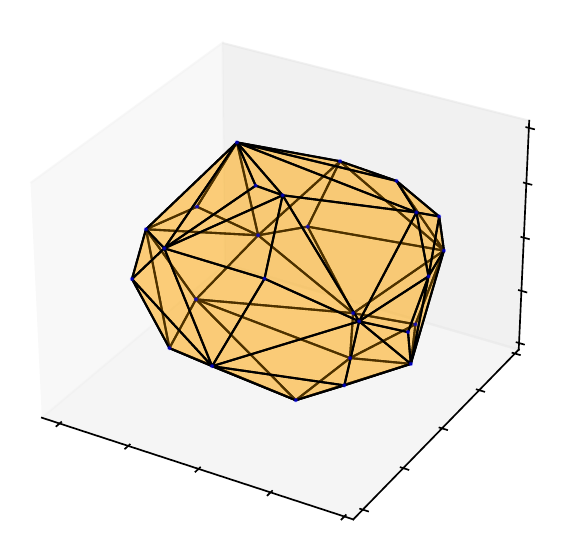}
    \caption{Phase 6}
    \label{fig:ray}
  \end{subfigure}
  \caption{Illustration of randomly distributed points (50 points), showcasing the polyhedron formed by the extreme (green) and corner (red) points. This figure summarizes all the phases of the algorithm.}
  \label{fig:algorithm}
\end{figure*}

The focus of this research is to design and implement a parallel preprocessing filter algorithm to speed up current 3D implementations of convex hull computation through the efficient use of CUDA, Tensor and Ray tracing cores. This section describes each step of the algorithm  presented in Algorithm \ref{algo:main} and illustrated  in Figure \ref{fig:algorithm}. The main algorithm consists of the following steps: (1) Finding axis  extreme points, (2) Finding the polyhedron corners, (3) Building the filtering polyhedron, (4) Filtering hull candidate points, (5) Compacting the candidate point set and, finally, (6) Computing the convex hull from candidate points. Steps 1-5 are executed in GPU (all functions that start with the word "Parallel" are executed in GPU) and step 6 in CPU.

\begin{algorithm*}[ht!]
    \caption{Filter main algorithm}
    \label{algo:main}
    \begin{algorithmic}[1]
    \Require Set of points $S$ 
    \Ensure Set of points to the hull 
    
    \Statex \textbf{Phase 1 and 2: Axis and Corner Analysis}
    \State $\{P^{min}_i, P^{max}_i\} \gets \Call{ParallelMinMaxReduction}{S}$ \Comment{Subsection \ref{s:extremepoints}}
    \State $B \gets \Call{ParallelGetBoundingBoxCorners}{P^{min}_x, P^{min}_y, P^{min}_z, P^{max}_x, P^{max}_y, P^{max}_z}$ \Comment{Subsection \ref{s:corner}}
    
    \Statex \textbf{Phase 3: Build Filtering Polyhedron}
    \State $Scene \gets BuildPolyhedron(\{P^{min}_x, P^{max}_x, P^{min}_y, P^{max}_y, P^{min}_z, P^{max}_z, C_1, \dots, C_8\})$ \Comment{Subsection \ref{s:filter}}
    
    \Statex \textbf{Phase 4: Ray-based Filtering}
    \State $FilteredPoints \gets \Call{GenerateRay}{MasterPoint, TargetPoints, Scene, Results}$ \Comment{Subsection \ref{s:ray}}
    
    \Statex \textbf{Phase 5: Output Compaction}
    \State $CompactedArray \gets \Call{Compaction}{FilteredPoints}$ \Comment{Subsection \ref{s:discard}}
    
    \Statex \textbf{Phase 6: Hull Computation}
    \State $Hull \gets \Call{ComputeConvexHull}{CompactedArray}$ \Comment{Subsection \ref{s:ch}}
    
    \State \Return $Hull$
    \end{algorithmic}
\end{algorithm*}

\subsection{Finding Axis Extreme Points}\label{s:extremepoints}

The goal is this step is to find extreme points, points that belong to the hull,   to build a  filtering polyhedron that allow us discard all the points that are inside this polyhedron. We start  finding  6 extreme points, the ones defined by the minimum and maximum values in each coordinate axis  ($min_x, max_x, min_y, max_y, min_z, max_z$) coordinates. To find these extreme points we used the min-max parallel reduction strategy  in each axis as shown in \cref{line:extremes} of Algorithm \ref{alg:findExtremesAndCorners}. This strategy works iteratively by comparing two array elements and selecting the minimum or the maximum values, respectively. After each iteration, the number of elements to be compared is reduced by half. The parallel min-max reduction operation is $O(\log n)$ in time complexity. One well-known algorithm for parallel reductions is the one presented by Harris et al \cite{harris_2005,harris2007optimizing}, which is presented in Algorithm \ref{alg:minmaxreduction}. 

\begin{algorithm*}[h!]
\caption{Finding Representative Polyhedron Points}
\label{alg:finding_polyhedron}
\begin{algorithmic}[1]
    \Require $Points$: Array of 3D vectors $\mathbf{v} = (x, y, z)$.
    \Ensure $\mathcal{S}$: Set of 14 extreme and corner-proximal points.
    
    \Statex \textbf{Step 1: Find Extreme Points per Axis}
    \For{$k \in \{x, y, z\}$}
        \State $P^{min}_k, P^{max}_k \gets \text{ParallelMinMaxReduction}(k, Points)$ \label{line:extremes}
    \EndFor
    
    \Statex \textbf{Step 2: Define Bounding Box Corners}
    \State $\mathcal{B} \gets \text{ParallelGetBoundingBoxCorners}(P^{min}_x, \dots, P^{max}_z)$ \Comment{$\mathcal{B} = \{B_1, \dots, B_8\}$}
    
    \Statex \textbf{Step 3: Find Nearest Points to Corners (Manhattan Distance)}
    \For{each corner $B_j \in \mathcal{B}$} \label{line:init_corners}
        \For{each point $v \in Points$ \textbf{in parallel}}
            \State $Distances[v] \gets \sum_{k \in \{x,y,z\}} |B_{j,k} - v_k|$ \label{line:manhattan}
        \EndFor
        \State $C_j \gets \text{ParallelMinPointReduction}(Distances, Points)$
    \EndFor \label{line:end_corners}
    
    \State \Return $\mathcal{S} \gets \{P^{min}_x, P^{max}_x, P^{min}_y, P^{max}_y, P^{min}_z, P^{max}_z, C_1, \dots, C_8\}$
\end{algorithmic}
\label{alg:findExtremesAndCorners}
\end{algorithm*}

\begin{algorithm*}[h!]
\caption{GPU Parallel Min/Max Reduction}
\label{alg:minmaxreduction}
\begin{algorithmic}[1]
    \Require $A$: Input array in global memory, $N$: Number of elements.
    \Ensure $R$: Array of partial results (one per block).
    
    \Statex \textbf{Shared Memory Allocation:}
    \State $sMin[T], sMax[T] \gets$ \text{Shared memory arrays of size block dimension}
    
    \Procedure{MinMaxReduction}{$A, R, N$}
        \State $tid \gets \text{threadIdx.x}, \quad bid \gets \text{blockIdx.x}$
        \State $gid \gets bid \times \text{blockDim.x} + tid$
        \State $step \gets \text{gridDim.x} \times \text{blockDim.x}$
        
        \State $min_{local} \gets +\infty, \quad max_{local} \gets -\infty$
        
        \Statex \Comment{Grid-stride loop to handle $N >$ total threads}
        \For{$i = gid$ \textbf{to} $N-1$ \textbf{step} $step$}
            \State $min_{local} \gets \min(min_{local}, A[i])$
            \State $max_{local} \gets \max(max_{local}, A[i])$
        \EndFor
        
        \State $sMin[tid] \gets min_{local}$
        \State $sMax[tid] \gets max_{local}$
        \State \textbf{syncthreads}() \Comment{Wait for all threads in block}
        
        \Statex \Comment{Tree-based reduction in shared memory}
        \For{$stride = \text{blockDim.x} / 2$ \textbf{down to} $1$ \textbf{step} $stride / 2$}
            \If{$tid < stride$}
                \State $sMin[tid] \gets \min(sMin[tid], sMin[tid + stride])$
                \State $sMax[tid] \gets \max(sMax[tid], sMax[tid + stride])$
            \EndIf
            \State \textbf{syncthreads}()
        \EndFor
        
        \Statex \Comment{Write block result to global memory}
        \If{$tid = 0$}
            \State $R[bid].min \gets sMin[0]$
            \State $R[bid].max \gets sMax[0]$
        \EndIf
    \EndProcedure
\end{algorithmic}
\end{algorithm*}


\subsection{Finding Polyhedron Corners} \label{s:corner}
The 6 extreme-points (green points from Figure \ref{fig:input}) found in Section \ref{s:extremepoints} allows building an eight triangle-face polyhedron. We know that a way 
to improve the filtering step, that is, to discard more points,  is to use a polyhedron with a volume larger than the mentioned 8 triangle-face polyhedron. A larger polyhedron  allows one then to decrease the number of  candidate points to the hull, and so reduce the time to compute the convex hull of a point set. That is possible adding more points to the polyhedron found in the previous stage.  We propose to obtain additional  extreme points  from points furthest from the extreme points obtained in the previous step (it is not necessary that they be the furthest, but one of the furthest). These points together with the axial ends, correspond to the vertices of the polyhedron in Figure \ref{fig:poly_filter}. Finding the additional  extreme points with respect to the faces of a polyhedron  can be a computationally expensive task, depending on the method employed. A simple strategy consists of  identifying  for each polyhedron face, the point that maximizes the volume of the tetrahedron formed by this point and the polyhedron face. For instance, this is achieved by selecting the leftmost point, the point that is furthest in front, and the point that is highest above the reference frame. Taking into account this combination of extreme points, a reasonable estimate of the maximum volume can be obtained.

However, the cost of the traditional method of calculating volume is high since it involves calculating multiple vector products and matrix points; fortunately, there are other much more convenient strategies that only use simple operations such as addition and subtraction, such as the Manhattan distance, which is the operation we use in this work to find the vertices of the polyhedron, as seen in step three of the Algorithm \ref{alg:findExtremesAndCorners} in \cref{line:manhattan} where the calculation of this operation only involves three unsigned subtractions and two additions. Between \crefrange{line:init_corners}{line:end_corners}, we can see how the point closest to the corner of the bounding box is obtained using a parallel reduction to find the point.

\subsection{Building the Filtering Polyhedron} \label{s:filter}

The filtering polyhedron is constructed from 14 points, comprising the 6 extreme  points along the axes and the 8 points (corners)  identified above, as shown in figure \ref{fig:poly_filter}. This polyhedron, constructed from these points, consists of 24 triangular faces (each previous face is replaced by 3 new ones). So the filter criterion is: any point located inside this polyhedron must be discarded, conversely, if a point is located outside, it is a potential candidate for the hull.

\begin{algorithm}[h!]
\caption{Ray Tracing Pipeline (RT Core Abstraction)}
\label{alg:raygen_rtx}
\begin{algorithmic}[1]
    \Require $\mathbf{P}_{target}$: Target master point, $\mathcal{V}$: Array of origin points, $Scene$: Acceleration structure.
    \Ensure $FilteredPoints$: Array containing hit data.

    \Procedure{RayGeneration}{} \label{line:init_raygen}
        \State $idx \gets \text{Global invocation index}$
        \State $\mathbf{O} \gets \mathcal{V}[idx]$ \Comment{Origin of the ray}
        \State $\mathbf{D} \gets (\mathbf{P}_{target} - \mathbf{O})$ \Comment{Direction vector}
        
        \State $t_{min}, t_{max} \gets 0.0, \infty$
        \State $payload \gets \text{MISS}$ \Comment{Initialize ray state, no hit 
        }

        \State \Call{LaunchRay}{$Scene, \mathbf{O}, \mathbf{D}, t_{min}, t_{max}, payload$}
        
        \State $FilteredPoints[idx] \gets payload$
    \EndProcedure \label{line:end_raygen}

    \Statex
    \Procedure{AnyHit}{$payload$} \label{line:init_hit}
        \State $payload \gets \text{HIT}$
        \State \Call{AcceptHitAndTerminate}{} 
    \EndProcedure \label{line:end_hit}

    \Statex
    \Procedure{MissHit}{$payload$} \label{line:init_misshit}
        \State $payload \gets \text{MISS}$
    \EndProcedure \label{line:end_misshit}
\end{algorithmic}
\end{algorithm}

\subsection{Filtering hull candidate points} \label{s:ray}

The main idea of the proposed algorithm is inspired by the fact that ray tracing cores can be used to emit rays and to check the hits with the faces of the modeled object. In this case the modeled object is the  polyhedron formed by the triangle faces built in the previous step,  and the query is find if a point $p$ is inside or outside this polyhedron. In order to do this, a ray is launched from each point $p$ outwards (outside of the polyhedron) as shown in Figure \ref{fig:ray}. If the ray hits a polyhedron face, it means that $p$ is inside and must be discarded. If the ray does not hit any face, it means it is a candidate point to the convex hull and is kept for the next step. 

To implement the previous algorithm using RT cores, three phases are employed: the first is the configuration phase, followed by the BVH construction phase, and finally the ray launch phase.

\subsubsection{Configuration phase}
In the configuration phase, all the meta parameters required by the Ray Tracing code are established, such as the number of rays to be used, which in this case is the same number as the input points, the size and characteristics of the BVH, the meta data of the functions that describe the behavior of the rays, and the characteristics and format of the inputs and output of the RT routines.

\subsubsection{BVH phase}
The BVH is the data structure used to contain the faces of the polyhedron, this data structure is built using the faces formed by the 14 extreme points. The time spent building the BVH is around $1.9$ millisecond for a constant amount of triangles (24 faces) and any amount of points to intersect as input, where it is possible to notice that the time has a very low variation for both distributions (described in section \ref{subsec:exp} and figure \ref{fig:dist}) with any number of input points, 
The BVH data structure is accelerated by hardware and constructed using the set of functions provided by OptiX.


\subsubsection{Ray/triangle Intersection phase}

Three behaviors of rays are required to implement the strategy:

\begin{itemize}
    \item Ray Generation:
The rays are generated using each point $p$ as origin and a direction defined as $p-q$, where $q$ is a point inside the polyhedron as seen in the figure \ref{fig:ray}. We use as point $q$  the midpoint on the x axis, but it could be any other point inside it. A distance between $0$ and $1M$ unit of distance must also be specified as the maximum final distance, where we use the maximum possible distance. This is exposed in \crefrange{line:init_raygen}{line:end_raygen} of Algorithm \ref{alg:raygen_rtx}.


    \item Any hit:
When a ray hits a face of the polyhedron, the point $p$  (ray origin) is marked as non-candidate to the hull. 

    
    \item Miss hit:
If a ray does not hit any side of the polyhedron, it means that it is a point that is possibly part of the hull and is marked in its corresponding location in the array called \textit{FilteredPoints} as a candidate.
\end{itemize}





\subsection{Compacting the candidate point set} \label{s:discard}

As a result of the previous phase, a \textit{FilteredPoints} array is obtained that indicates which points can be discarded with complete certainty, since they cannot be part of the hull because they are inside the filtering polyhedron in the algorithm \ref{alg:compaction}. However, the input remains the same size as at the beginning, since an array of 0 and 1 is obtained, where 0 indicates that it does not belong and 1 indicates that it is a candidate. This phase builds a candidate point set from the input point set by keeping only  the candidate points for the hull.

The input point set array can be seen as an sparse array of candidate points. Reducing the length of a sparse array on a GPU is challenging, as parallel operations on each array element are unaware of the new positions in a compacted array. However, there are techniques that rely on the prefix addition algorithm to determine which positions should be moved. In this work, we employ a three-level compaction approach (algorithm \ref{alg:tc_scan_full}) that takes advantage of new technologies in modern GPUs, using tensor kernels based on \cite{ScanTC,9147055} to perform prefix sum computations where specific optimization was performed on the kernel hierarchy and modifications to work in integer precision for this problem. At the first level (algorithm \ref{alg:tc_scan_full}, \cref{line:scanTC}), obtaining as a result the cumulative sum of the blocks corresponding to the bulk of the array to be compacted by TC with MMA options (algorithm \ref{alg:tc_scan_full}, \cref{line:MMA}) on the matrix previously configured to obtain the sum of prefixes of group of 256 elements by warp, then we work with a kernel hierarchy at the warp and block level respectively (algorithm \ref{alg:tc_scan_full}, \crefrange{line:WarpScan}{line:blockScan}) to obtain the cumulative sums. Later levels employ traditional techniques such as inclusive cub scanning by cub library (algorithm \ref{alg:tc_scan_full}, \cref{line:scanCUB}) to obtain the cumulative total of all blocks, and the last level of the hierarchy is a top-down sweep kernel (algorithm \ref{alg:tc_scan_full} to sum the cumulative total of all levels of the block hierarchy, warps and theads, \cref{line:scanSM}). Finally, the point positions are re-assigned in parallel according to their cumulative sum relative positions.

\begin{algorithm}[h!]
\caption{Parallel Stream Compaction for Polyhedron Candidates}
\label{alg:compaction}
\begin{algorithmic}[1]
    \Require $Points$: Original array of $N$ points, $FilteredPoints$: Boolean flags (1 if candidate, 0 otherwise).
    \Ensure $Compacted$: Array containing only points where $FilteredPoints[i] = 1$.

    \Procedure{Compaction}{$Points, FilteredPoints, N$}
        \State $Indices \gets \text{TCParallelScan}(FilteredPoints)$ 
        \Statex \Comment{Compute scatter addresses}
        \State $Total \gets Indices[N-1] $ \Comment{Size of the output array}
        \State $Compacted \gets \text{Allocate array of size } Total$
        
        \For{$i = 0$ \textbf{to} $N-1$ \textbf{in parallel}}
            \If{$Filtered[i] = 1$}
                \State $dest \gets Indices[i]$
                \State $Compacted[dest] \gets Points[i]$
            \EndIf
        \EndFor
        
        \State \Return $Compacted$
    \EndProcedure
\end{algorithmic}
\end{algorithm}

\begin{algorithm*}[h!]
\caption{Tensor Core Accelerated Scan (Host \& Device)}
\label{alg:tc_scan_full}
\begin{algorithmic}[1]
    \Require $In$: Input array, $N$: Number of elements.
    \Ensure $Out$: Output array with inclusive prefix sum.

    \Statex \Comment{\textbf{--- Part 1: Host Code (Main Program) ---}}
    \Procedure{TCParallelScan}{$In$}
        \State $NumSegments \gets \lceil N / 256 \rceil$ \Comment{Small tiles processed by Tensor Cores}
        \State $NumBlocks \gets \lceil N / 8192 \rceil$   \Comment{Large blocks (32 segments per block)}
        
        \State \textbf{Allocate} $SumsBlock, SumsWarp, SumsThread$
        
        \State \Comment{Phase 1: Launch TC Kernel (Local Scan)}
        \State \Call{TCScanKernel}{$SumsThread, SumsWarp, SumsBlock, In$} \label{line:scanTC}
        
        \State \Comment{Phase 2: Scan Block Accumulators (Global Connectivity)}
        \State $Aux \gets \Call{CUBScan}{SumsBlock}$  \label{line:scanCUB}
        
        \State \Comment{Phase 3: Downsweep / Add Offsets}
        \State \Call{AddPartialSums}{
        ($Out, SumsThread, SumsWarp, SumsBlock, Aux$)} \label{line:scanSM}

        \State \Return $Out$
        
    \EndProcedure

    \Statex \hrulefill
    \Statex \Comment{\textbf{--- Part 2: Device Code (TC Kernel) ---}}
    
\Procedure{TCScanKernel}{$D_{thread}, S_{warp}, S_{block}, D_{in}$}
        \State $tid \gets \text{threadIdx.x}$
        \State $LaneID \gets tid \pmod{32}$
        
        \State \Comment{Calculate IDs for memory addressing}
        \State $LocalWarpID \gets tid / 32$
        \State $GlobalWarpID \gets (tid + \text{blockDim.x} \times \text{blockIdx.x}) / 32$
        \State $Offset \gets (LocalWarpID \times 256) + (\text{blockIdx.x} \times \text{SEGMENT\_SIZE})$

        \State \Comment{\textbf{Step A: Tensor Core Scan}}
        \State \Call{MMAScan}{$D_{in}, D_{out}, \dots$}  \label{line:MMA}
        \Comment{Performs scan on 256 elements by Warp}

        \State \Comment{\textbf{Step B: Collect Partial Sums per Warp}}
        \label{line:WarpScan}
        \State $WarpsPerBlock \gets \text{BlockDim.x} / 32$

        \If{$tid < WarpsPerBlock$}
            \State $acc \gets D_{thread}[idx*LaneID]$ \Comment{Get total sum of this warp's tile}
        \EndIf
        
        \State \Comment{\textbf{Step C: Intra-Block Warp Scan}}
        \State $acc \gets \Call{WarpInclusiveScan}{acc}$
        \label{line:WarpScan}

        \State \Comment{\textbf{Step D: Store Hierarchy Results}}
        \If{$tid < WarpsPerBlock - 1$}
            \State $S_{warp}[\text{GlobalWarpIdx}] \gets acc$ \Comment{Store total sum of the warp}
        \EndIf

        \If{$tid = WarpsPerBlock - 1$}
            \State $S_{block}[\text{blockIdx.x}] \gets acc$ 
            \label{line:blockScan}
            \Comment{Store total sum of the block}
        \EndIf

        
    \EndProcedure
    
    \Statex \hrulefill
    
\end{algorithmic}
\end{algorithm*}

\subsection{Computing the convex hull  from candidate points} \label{s:ch}

Finally, the proposed algorithm returns the candidate point set with less  less than or equal to the input set. The candidate point set can be used as input of   any state-of-the-art algorithm  to compute the convex hull. During the performance evaluation of the proposed strategy, we used the Pseudohull algorithm from the ParGeo library \cite{ParGeo}  to compute the convex hull, which is a highly competitive multi-core implementation.

\section{Experimental Evaluation}
\label{sec:experiment}

In this section, we conducted several experiments to evaluate  the performance and power consumption of the proposed GPU filter algorithm called \texttt{RTX} against \texttt{Pseudohull}, the state of the art  multicore filter algorithm. Both filter algorithms are used as preprocessing step of a traditional convex hull algorithm. First, section \ref{subsec:exp} describes the point distributions used  to evaluate   three implementations named as \texttt{RTX}, \texttt{Pseudohull} and \texttt{CUDA}. Second,  section \ref{subsec:exp_cpu}  presents  an evaluation of \texttt{Pseudohull} by considering the number and type of  CPU cores. Third,  section \ref{subsec:perfomance}, presents  a comparison of  \texttt{RTX}, \texttt{Pseudohull} \texttt{and} \texttt{CUDA}  implementations of the filter phase. Fourth, section \ref{subsec:ch} compares the performance of combining  \texttt{RTX} and \texttt{CUDA}  filter implementations with  \texttt{Pseudohull} (phase 1 and 2) to compute the convex hull.  Fifth,  section  \ref{subsec:power} shows the power consumption (Sub) of the proposed GPU implementation. Finally, in section \ref{subsec:future} we conclude this section with a demonstration of the performance of the proposed algorithm on different GPU architectures and a prediction of how it will continue to scale on future architectures (Subsection ).



All experiments were executed on the Yeco server, which is provided by the University of Chile Meshing for Applied Science Lab. The server has an Intel Core Ultra 7 265K processor with a maximum power consumption of 250 Watts and an average of 125 Watts. This processor offers two types of cores: 8 dedicated to high-performance processing at 3.9 GHz and 12 to low-power processing at 3.3 GHz, for a total of 20 cores. This machine has 32 GB of ram. It also features a NVIDIA RTX 4090 GPU (Ada Lovelace architecture, 2022) with 24 GB of graphics memory and 450 Watts of peak power consumption. The implementation is  compiled in C++ 13.3.0 and utilize the -O3 optimization level for the CPU. Additionally, for GPU processing, the code is implemented in CUDA with NVCC version 12.0, and single-precision floating-point arithmetic (FP32) is used for the  input points.

\subsection{Evaluation scenarios}
\label{subsec:exp}
\begin{figure*}[h!] 
    \centering
    \begin{subfigure}{0.49\textwidth}
        \centering
        \includegraphics[width=0.75\linewidth]{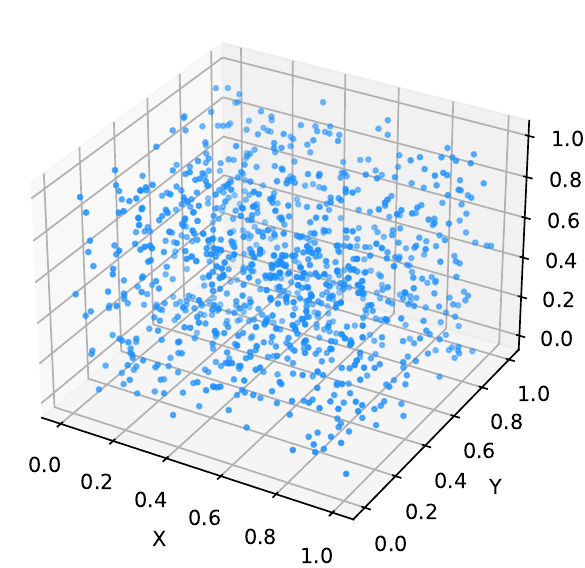}
        \caption{Uniform Distribution}
        \label{fig:uniform}
    \end{subfigure}
    \hfill
    \begin{subfigure}{0.49\textwidth}
        \centering
        \includegraphics[width=0.75\linewidth]{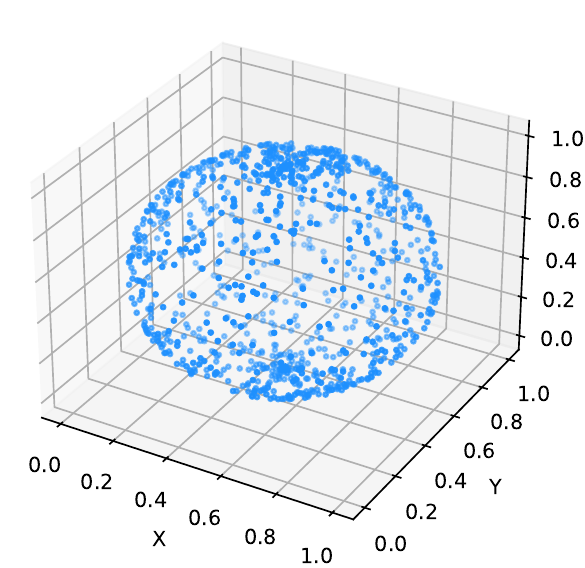}
        \caption{Sphere Distribution}
        \label{fig:sphere}
    \end{subfigure}
    
    \vfill
    \begin{subfigure}{0.49\textwidth}
        \centering
        \includegraphics[width=0.75\linewidth]{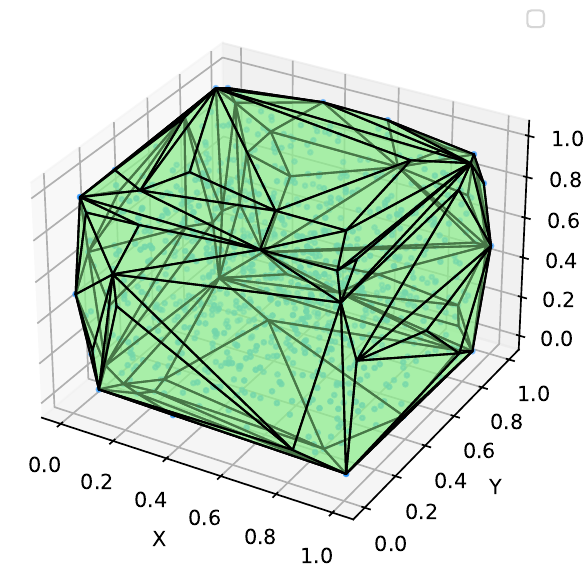}
        \caption{Convex Hull for Uniform Distribution}
        \label{fig:uniform}
    \end{subfigure}
    \hfill
    \begin{subfigure}{0.49\textwidth}
        \centering
        \includegraphics[width=0.75\linewidth]{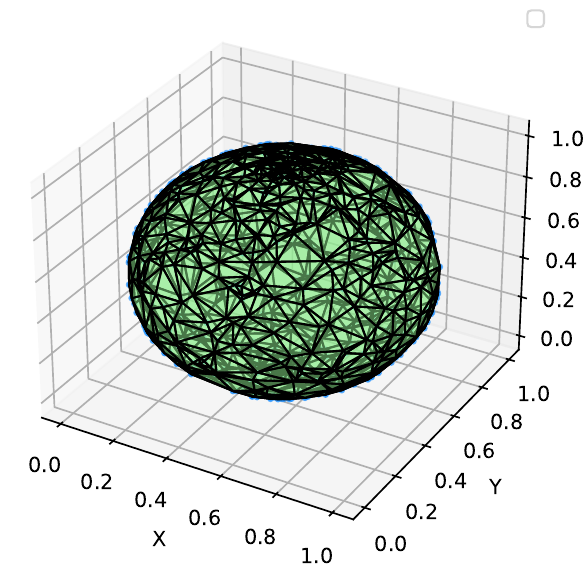}
        \caption{Convex Hull for Sphere Distribution}
        \label{fig:sphere}
    \end{subfigure}
    \caption{Randomly distributed points (1000 points) for an uniform and sphere distribution and their convex hulls respectively.}
    \label{fig:dist}
\end{figure*}

\begin{figure*}[htb!] 
    \centering
    \begin{subfigure}{0.3\textwidth}
        \centering
        \includegraphics[width=0.9\linewidth]{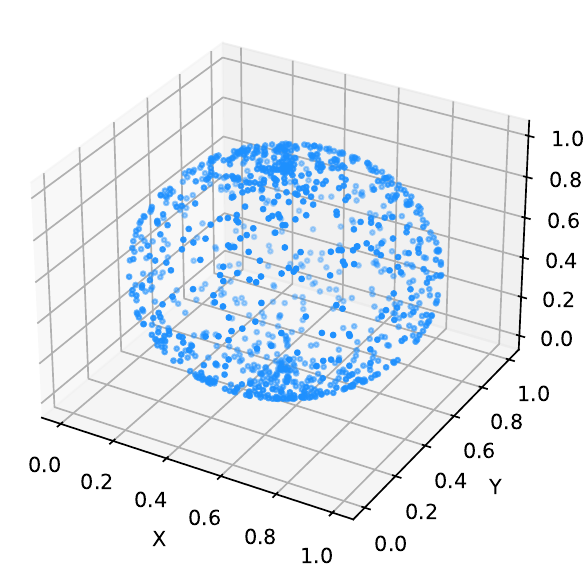}
        \caption{$\rho = 0.0$}
    \end{subfigure}
    \begin{subfigure}{0.3\textwidth}
        \centering
        \includegraphics[width=0.9\linewidth]{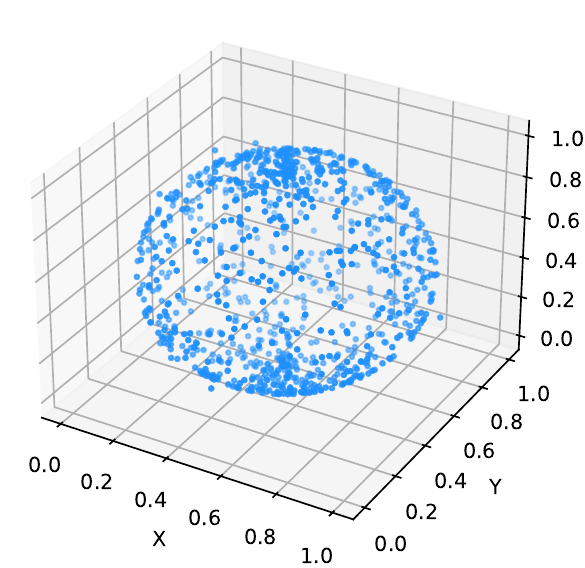}
        \caption{$\rho = 0.1$}
    \end{subfigure}
    \begin{subfigure}{0.3\textwidth}
        \centering
        \includegraphics[width=0.9\linewidth]{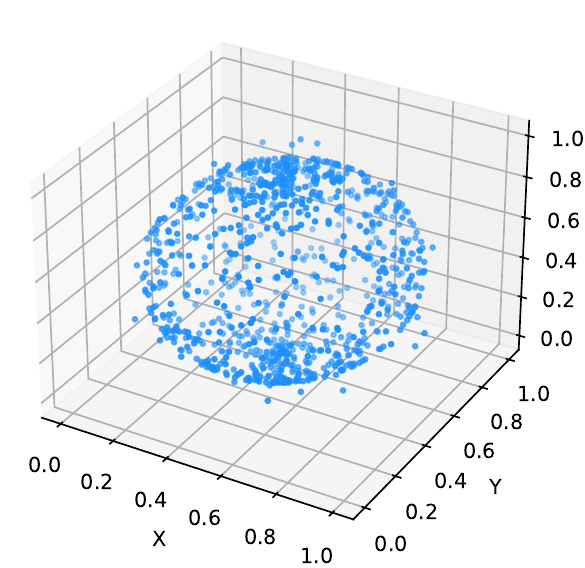}
        \caption{$\rho = 0.2$}
    \end{subfigure}

    \begin{subfigure}{0.3\textwidth}
        \centering
        \includegraphics[width=0.9\linewidth]{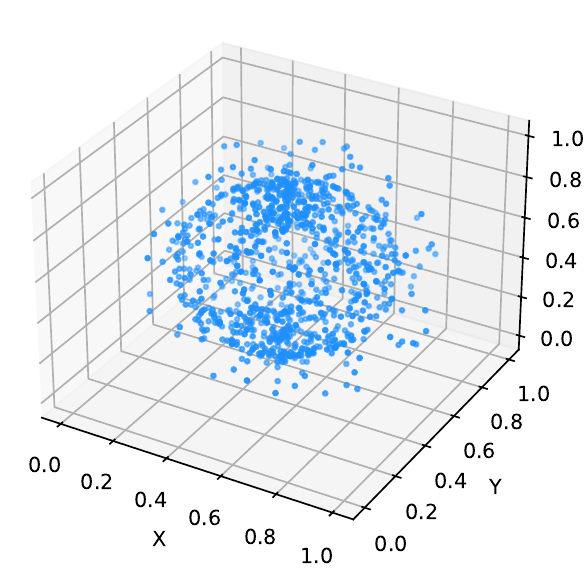}
        \caption{$\rho = 0.5$}
    \end{subfigure}
    \begin{subfigure}{0.3\textwidth}
        \centering
        \includegraphics[width=0.9\linewidth]{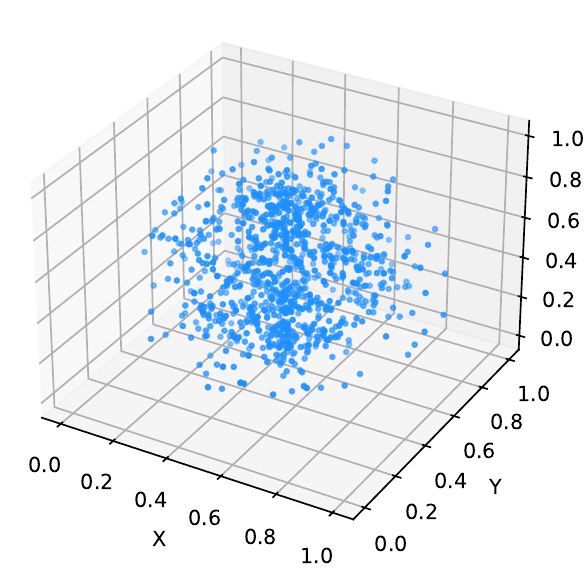}
        \caption{$\rho = 0.7$}
    \end{subfigure}
    \begin{subfigure}{0.3\textwidth}
        \centering
        \includegraphics[width=0.9\linewidth]{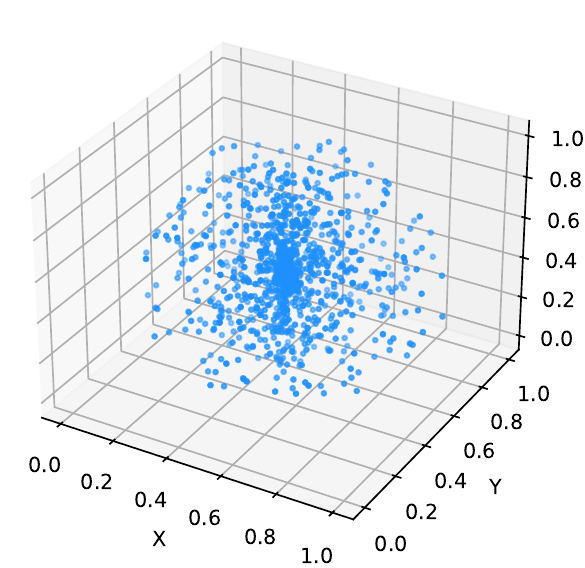}
        \caption{$\rho = 1.0$}
    \end{subfigure}
    \caption{Randomly sphere distribution points (1000 points) varying the $\rho$ value between $0$ to $1$.}
    \label{fig:sphere_dist}
\end{figure*}

Figure \ref{fig:dist} shows the two point distributions used in this work. The first distribution (uniform) is a scenario where a priori it is possible  to discard a large number of points with a filtering phase and the second point distribution, points are located on the sphere  with a displacement controlled by the a noise parameter called $\rho$ as shown in  Figure \ref{fig:sphere_dist}.  It is possible to change the thickness of the surface according to the rate $\rho$ between 0 and 1, where 0 mean all points are on the surface of the sphere (the worst case because all input points are candidate points) and 1 mean all points are inside of the sphere. In this manner, we can control the number of points that will be discarded during the filtering phase and measure from which $\rho$ it is advantageous to use a filter.

\begin{table}[h!]
\begin{tabular}{ll}
    \# Variant       & Type of Parallelism           \\
\hline
    RTX             & GPU: CUDA, RT and Tensor cores               \\
    CUDA            & GPU: CUDA and Tensor cores                   \\
    Pseudohull        & CPU: performance and energy efficient cores               
\end{tabular}
\caption{This table shows all the variant used in the experimentation evaluation using GPU and CPU and the type of parallelism employed.}
\label{tab:exp}
\end{table}

Table \ref{tab:exp} summarizes the three strategies to be evaluated. The first one, the \texttt{CUDA} implementation which benefits from CUDA and tensor cores, the \texttt{RTX} implementation that in addition to the previous cores,  benefits from  ray tracing cores and, the \texttt{pseudohull} implementation, provided by the Pargeo library, which benefits from a multi-core architecture.

All input point sets were randomly generated, using standard C++ libraries, where the uniform distribution is generated in the range [0,1]. On the other hand, the sphere is centered in [0.5, 0.5] and radius 0.5. It is also important to highlight that in these experiments we use between ten to sixteen different seeds for each input size depending of the experiment, in addition, we repeat each run of the algorithm 20-100 times for each seed, in order to reduce the dispersion of each measurement.


\subsection{Performance comparison of the filter algorithms}
\label{subsec:perfomance}
\begin{figure*}[htb!] 
    \centering
    \begin{subfigure}{0.49\textwidth}
        \centering
        \includegraphics[width=1\linewidth]{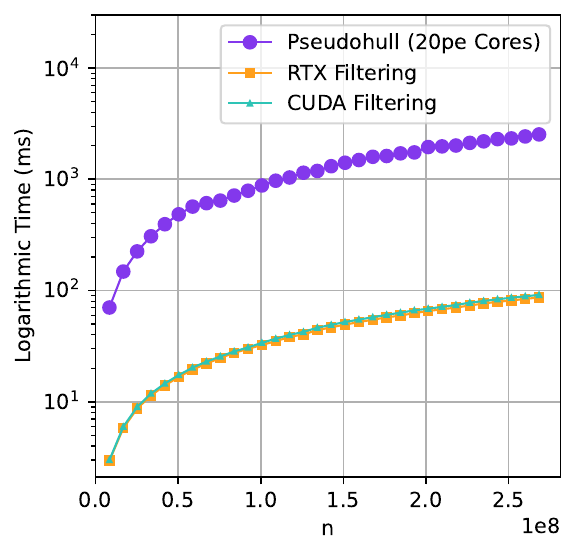}
        \caption{Uniform distribution}
    \end{subfigure}
    \hfill
    \begin{subfigure}{0.49\textwidth}
        \centering
        \includegraphics[width=1\linewidth]{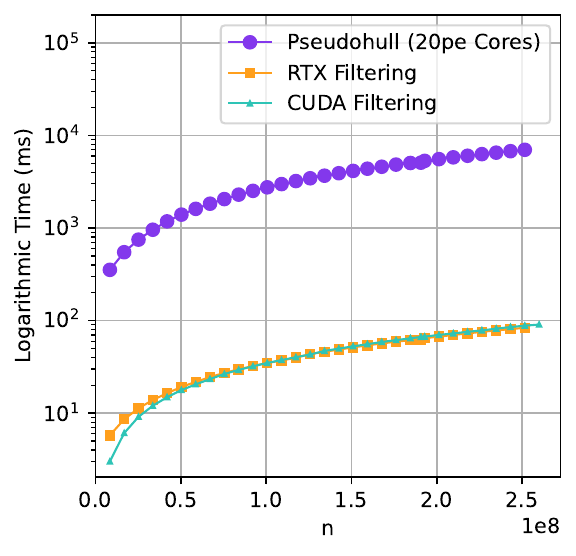}
        \caption{Sphere distribution}
    \end{subfigure}
    \caption{ Filtering time, where y-axis is in logarithmic scale, for the RTX filtering, CUDA filtering, and fastest CPU implementation (Pseudohull with 20 cores) in both distributions.}
    \label{fig:log_time}
\end{figure*}

This experiment measures the filter performance of the proposed GPU implementations (\texttt{CUDA} and \texttt{RTX}) and \texttt{Pseudohull} filter (first phase) with 20 cores, described in Table \ref{tab:exp}, on the two mentioned  distributions in the range of $2^{23}$ to $2^{28}$ points using uniform and sphere distributions. Figure \ref{fig:log_time} shows that the fastest variant is the \texttt{RTX}, followed by the \texttt{CUDA}.

However, the \texttt{CUDA} implementation is faster than the \texttt{RTX} implementation  when the filtering polyhedron is small enough, since the main optimization of this implementation is that no BVH is built to compute each ray, this allows us to ask if each ray hits a side of the polyhedron using a simple loop, which is fast for small polyhedron (24 in this case), however, the processing cost increases significantly when scaling the solution.

\begin{figure*}[htb!] 
    \centering
    \begin{subfigure}{0.49\textwidth}
        \centering
        \includegraphics[width=.98\linewidth]{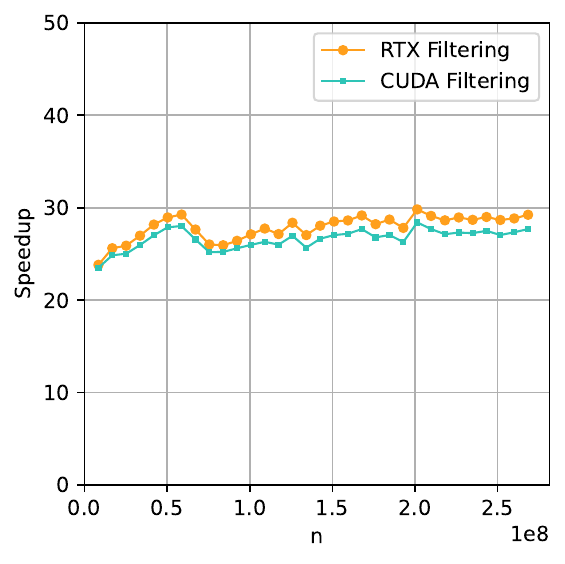}
        \caption{Uniform distribution}
    \end{subfigure}
    \hfill
    \begin{subfigure}{0.49\textwidth}
        \centering
        \includegraphics[width=1\linewidth]{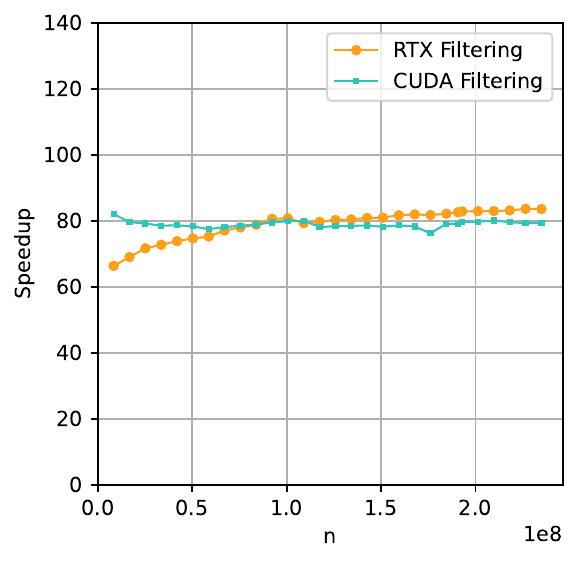}
        \caption{Sphere distribution}
    \end{subfigure}
    \caption{
Filtering speedup for the \texttt{RTX}, \texttt{CUDA}, over the fastest CPU implementation (\texttt{Pseudohull} with 20 cores) in both distributions.}
    \label{fig:speedup}
\end{figure*}

On the other hand, Figure \ref{fig:speedup} shows a speedup with a 20-core \texttt{Pseudohull} filter, where the RT implementation is slightly faster than the \texttt{CUDA} cores implementation, reaching between $24 \sim 30\times$ for a uniform distribution and $65 \sim 83 \times$ for a sphere distribution. Importantly, for the sphere distribution, the \texttt{RTX} implementation shows better scalability for a large number of points, due to the excellent ray manager in a large-scale BVH structure.

\subsection{Whole Convex Hull Performance}
\label{subsec:ch}
\begin{figure*}[htb!] 
    \centering
    \begin{subfigure}{0.49\textwidth}
        \centering
        \includegraphics[width=1\linewidth]{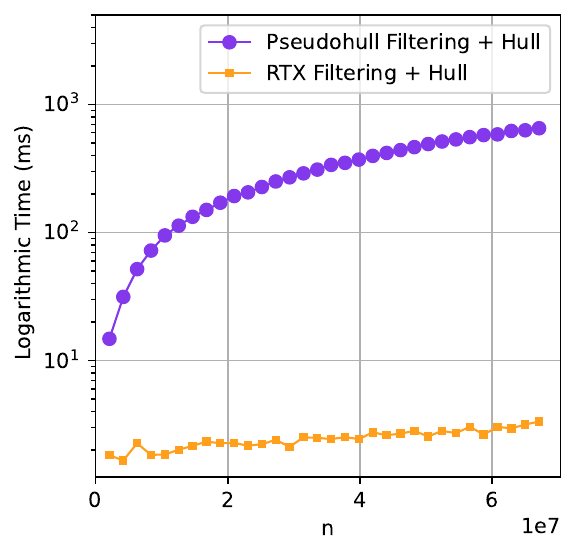}
        \caption{Uniform distribution}
    \end{subfigure}
    \hfill
    \begin{subfigure}{0.49\textwidth}
        \centering
        \includegraphics[width=1\linewidth]{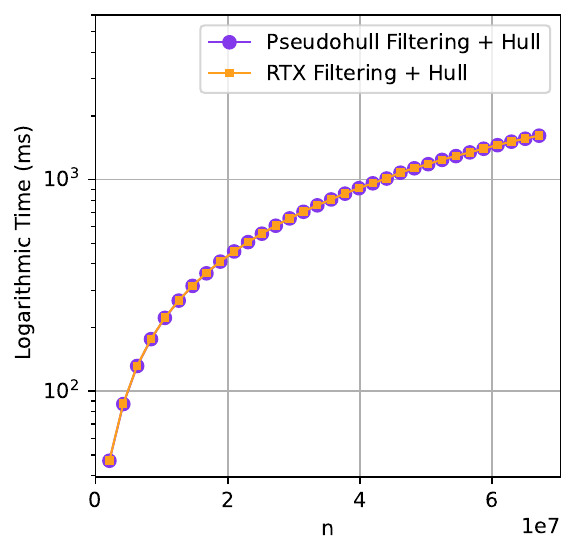}
        \caption{Sphere distribution}
    \end{subfigure}
    \caption{Convex hull time using the \texttt{RTX} filter, and \texttt{Pseudohull} algorithm with 20 cores in both distributions.}
    \label{fig:time_hull}
\end{figure*}

This experiment measures the runtime performance of the proposed filters for computing the convex hull, using the Pargeo library as a state-of-the-art benchmark. Figure \ref{fig:time_hull} shows the performance of the full hull algorithm on its own and using the filter, where both variants use \texttt{Pseudohull} to compute the full hull. It is possible to notice a notable difference between both distributions: in the uniform distribution, the variant that uses the RT-core filter is faster than the CPU-based filtering. However, in a sphere distribution, their performance is practically the same. This reflects the fact that it was not possible to filter out any points, but the filter does not significantly increase the overall time of the convex hull algorithm.

\begin{figure*}[h!] 
    \centering
    \begin{subfigure}{0.49\textwidth}
        \centering
        \includegraphics[width=0.99\linewidth]{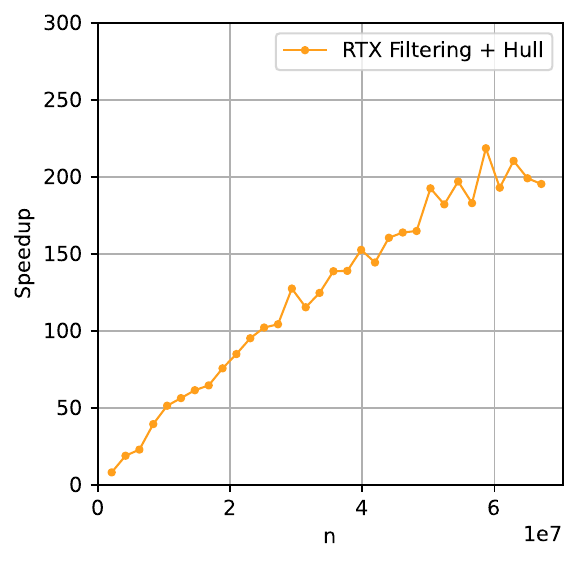}
        \caption{Uniform distribution}
    \end{subfigure}
    \hfill
    \begin{subfigure}{0.49\textwidth}
        \centering
        \includegraphics[width=1\linewidth]{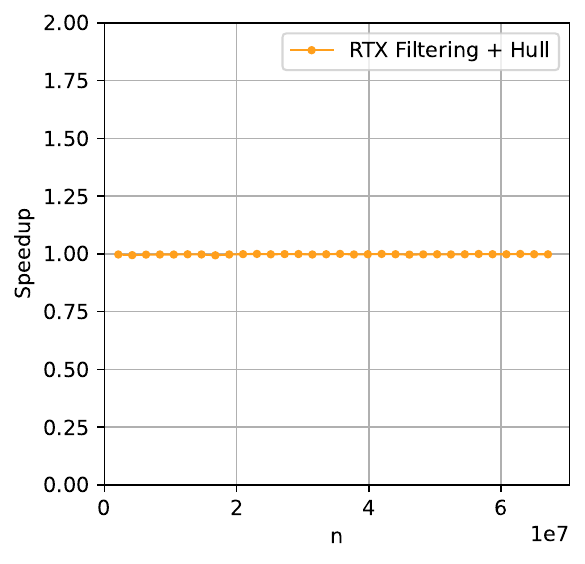}
        \caption{Sphere distribution}
    \end{subfigure}
    \caption{Convex hull speedup using the \texttt{RTX} filter over \texttt{Pseudohull} algorithm with 20 cores in both distributions.}
    \label{fig:log_hullspeedup}
\end{figure*}

Figure \ref{fig:log_hullspeedup} shows the performance improvement of the filter compared to the 20-core CPU variant, where it is approximately $210 \times$ faster than \texttt{pseudohull} variant for the uniform distribution. For the sphere distribution it shows a speedup of approximately $\sim1\times$, which is a positive result as well, as the input of the algorithm is not reduced by the filtering process neither was its performance.

\begin{figure}[h!] 
    \centering
    \includegraphics[width=1\linewidth]{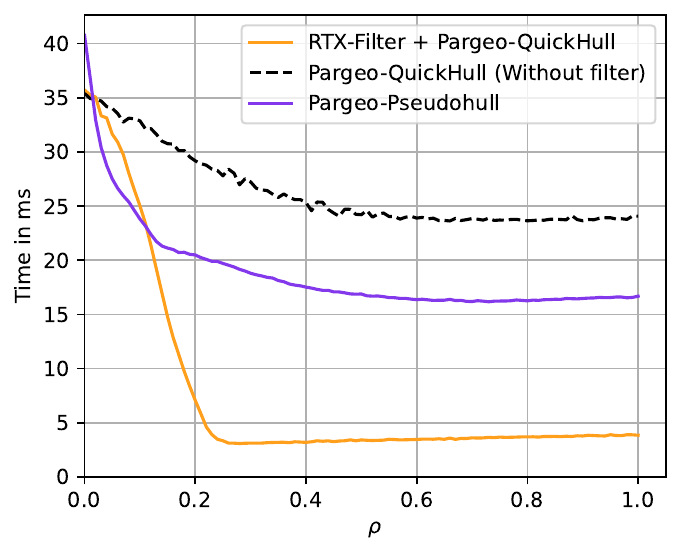}
    \caption{ Convex hull time varying $\rho$ value in a sphere for the \texttt{RTX} filter, \texttt{Pseudohull} and \texttt{QuickHull} provided by Pargeo both with 20 cores.}
    \label{fig:rho}
\end{figure}

It is possible to determine the number of points that need to be filtered to achieve faster processing speed. This is accomplished by moving the points within a sphere, as shown in Figure \ref{fig:sphere_dist}, using the variable $\rho \in [0,1]$ where $\rho = 0$ means all points are on the  sphere and $\rho=1$ means that all points are inside the sphere. Figure \ref{fig:rho} shows when a filtering algorithm is faster than a \texttt{Quickhull} algorithm alone provided by Pargeo Library. We can see that at $\rho=0.01$ the filter is faster than the non-filter algorithm, at $\rho=0.17$ is faster than the CPU filter, and $\rho \sim 0.25$ is the range of optimal performance.

\subsection{Power Consumption}
\label{subsec:power}
\begin{figure*}[htb!] 
    \centering
    \begin{subfigure}{0.49\textwidth}
        \centering
        \includegraphics[width=1\linewidth]{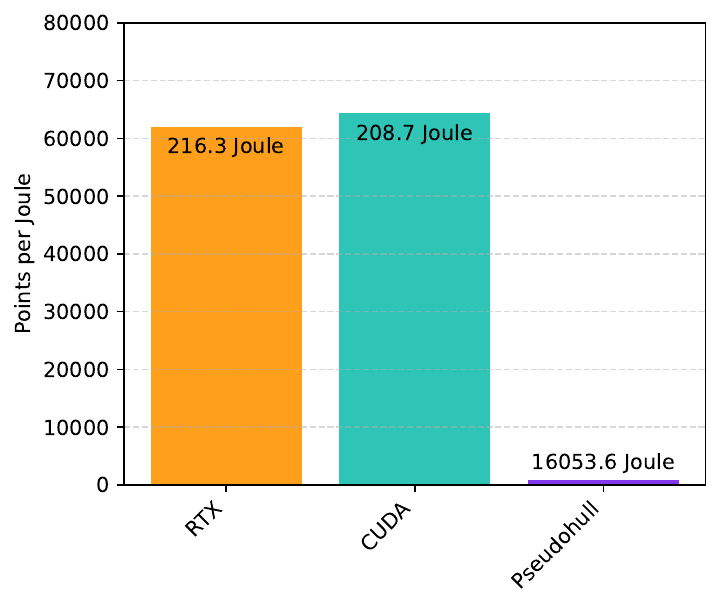}
        \caption{Uniform distribution}
        \label{fig:acc_energy_hull_uniform}
    \end{subfigure}
    \hfill
    \begin{subfigure}{0.49\textwidth}
        \centering
        \includegraphics[width=0.96\linewidth]{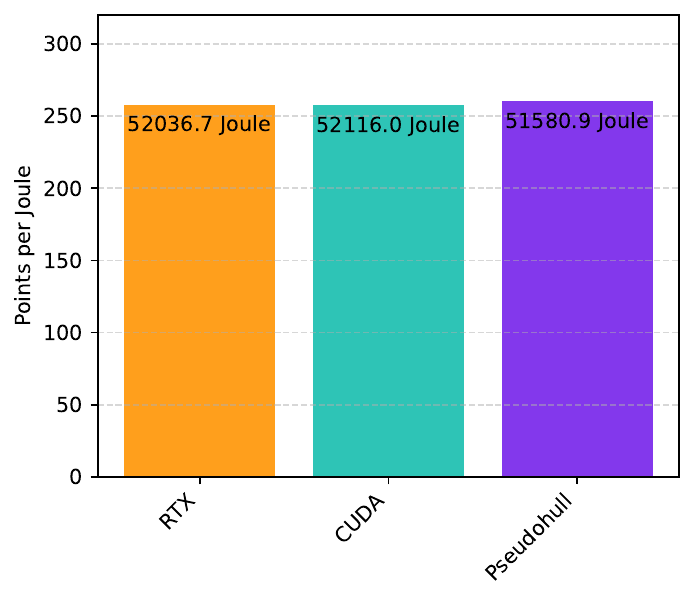}
        \caption{Sphere distribution}
        \label{fig:acc_energy_hull_sphere}
    \end{subfigure}
    \caption{Points per Joule (height) and accumulated energy (bar label) of the complete hull algorithm including the filter phase, in Joule  used for the GPU and CPU (\texttt{Pseudohull} with 20 cpu-cores) implementations for both distributions.}
    \label{fig:energy_hull}
\end{figure*}

Currently, the amount of energy used in data processing has become crucial for many governments and institutions worldwide. This experiment demonstrates how modern CPUs and GPUs use a low-cost energy algorithm to solve the convex hull. Figure \ref{fig:energy_hull} shows the energy consumption in joules for both distributions, where both GPU variants use $75\times$ less energy for the uniform distribution (Figure \ref{fig:acc_energy_hull_uniform}). Here, we observe that both GPU implementations do $\sim62,000$ $\text{Points}/\text{Joule}$  in $\sim210$ Joule, while the CPU implementation does $\sim830$ $\text{Points}/\text{Joule}$  in $\sim16,000$ Joule for the uniform distribution. However, for the sphere distribution, all implementations have a radical increase, reaching the highest consumption of $\sim52,000$ Joule and $\sim260$ $\text{Points}/\text{Joule}$ (Figure \ref{fig:acc_energy_hull_sphere}), this is due to the $99.5\%$ of time and energy used by the convex hull algorithm and only $0.5\%$ is the filter. 


\begin{figure*}[htb!] 
    \centering
    \begin{subfigure}{0.49\textwidth}
        \centering
        \includegraphics[width=1\linewidth]{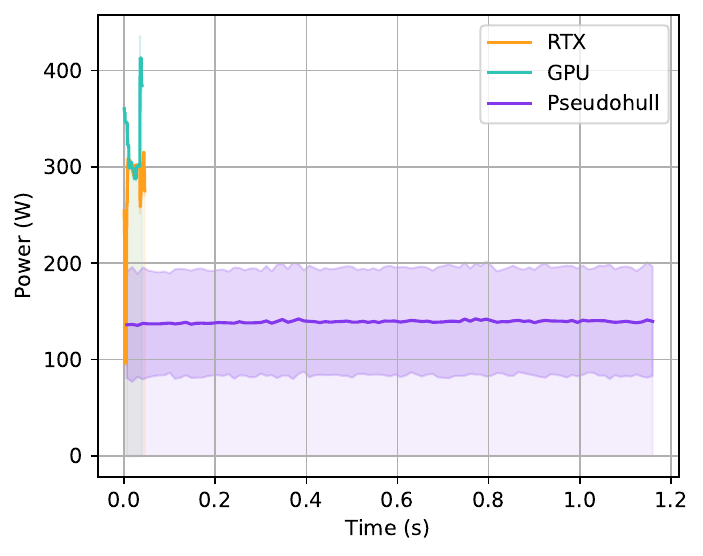}
        \caption{Uniform distribution}
        \label{fig:power_hull_uniform}
    \end{subfigure}
    \hfill
    \begin{subfigure}{0.49\textwidth}
        \centering
        \includegraphics[width=0.99\linewidth]{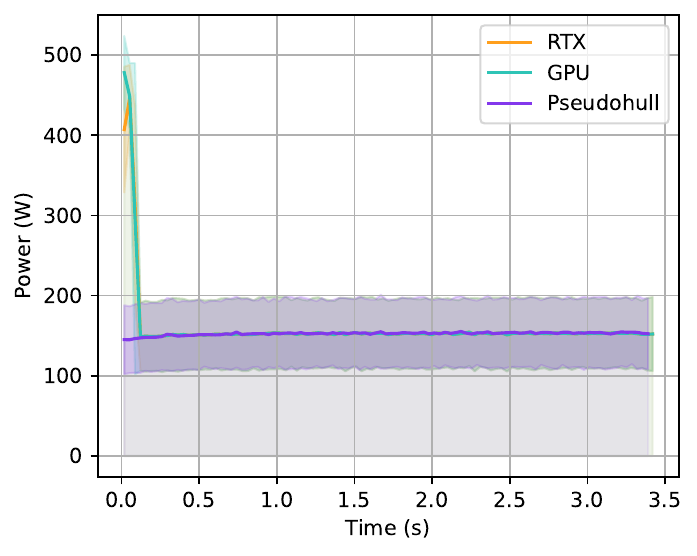}
        \caption{Sphere distribution}
        \label{fig:power_hull_sphere}
    \end{subfigure}
    \caption{Power in watts used for the entire full hull algorithm, including the filter phase where the shaded area is the deviation and the area under the curve is the consumption, used for GPU and CPU implementations (\texttt{Pseudohull} with 20 CPU cores) for both distributions.}
    \label{fig:power_hull}
\end{figure*}

On the other hand, the average power consumption for the uniform distribution is $\sim280$ Watts for \texttt{RTX}, $\sim320$ Watts for \texttt{CUDA}, and $\sim140$ Watts for \texttt{Pseudohull}. However, both GPU variants use more watts in a very short period of time, as shown in Figure \ref{fig:power_hull_uniform}. Unlike the CPU implementation, which is prolonged over time, this can also be observed in the sphere distribution, where all implementations are prolonged over time (see Figure \ref{fig:power_hull_sphere}). Since the hull algorithm is executed on the CPU, this gives an average power consumption of $\sim160$ Watts for both GPU implementations and $\sim150$ Watts for \texttt{Pseudohull}. This difference of $\sim10$ Watts is because we have added the idle power consumption for \texttt{RTX} and \texttt{CUDA}.

\subsection{Scalability}
\begin{figure*}[ht!] 
    \centering
        \centering
        \includegraphics[width=1
        \linewidth]{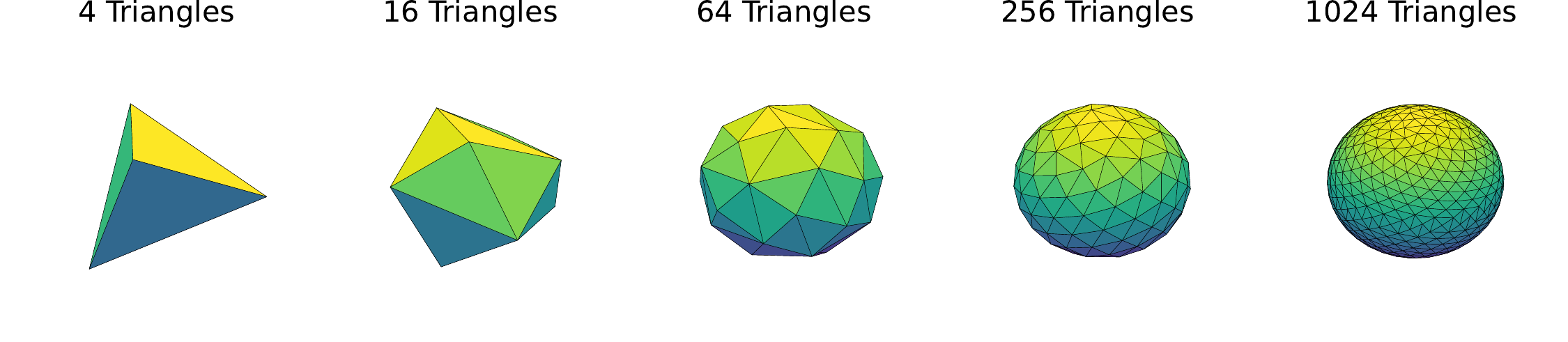}
        \caption{Illustration with the filter polyhedron of different numbers of triangles (faces) used in the experimentation..}
        \label{fig:scaling_poly_fig}
\end{figure*}

The main difference between \texttt{RTX}  and \texttt{CUDA} implementations is that the \texttt{RTX} implementation uses of BVHs and the RT cores while the  \texttt{CUDA} implementation   does not use them; therefore, it is much more efficient when the number of faces (triangles to intersect) and rays is smaller. Conversely, the variant with ray tracing cores is much more efficient when scaling to a large number of points intersecting the triangles, since the construction of the BVHs is also hardware-optimized. We tested this by measuring the execution time while varying the number of faces that make up the filtering polyhedron (as shown in Figure \ref{fig:scaling_poly_fig}) where the BVH building time and configuration of the RT cores are considered for the implementation of the \texttt{RTX}. However, the construction of a more robust filtering polyhedron considerably increases execution time and it is not taken in this experimentation. It is important to note that, on the left side of the plot, the difference between the both variant is less than 1 millisecond, while on the right side, the difference is exponential due to the logarithmic scale of the y-axis.

\begin{figure}[h!] 
    \centering
        \centering
        \includegraphics[width=1
        \linewidth]{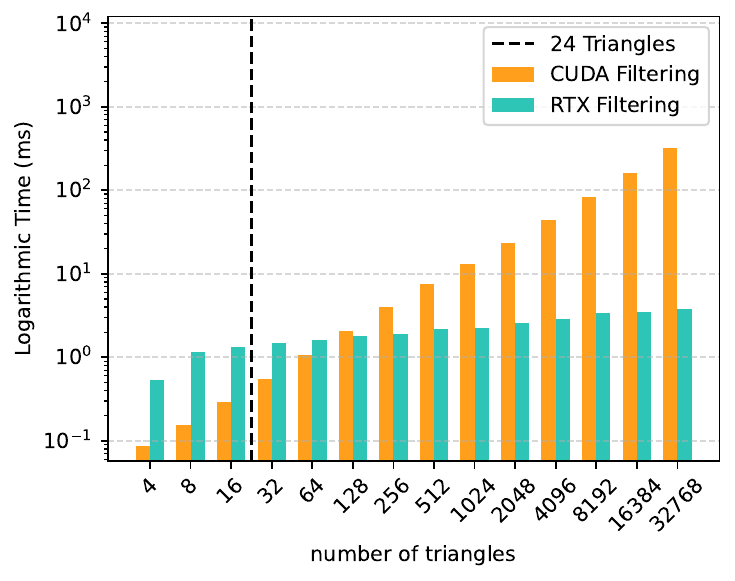}
        \caption{Bar chart for intersecting 33MM of points with x-axis representing the size of the filtered polyhedron and y-axis representing the time it takes for the triangles to intersect.}
        \label{fig:scaling_poly}
\end{figure}

\begin{figure}[htb!] 
    \centering
        \centering
        \includegraphics[width=1\linewidth]{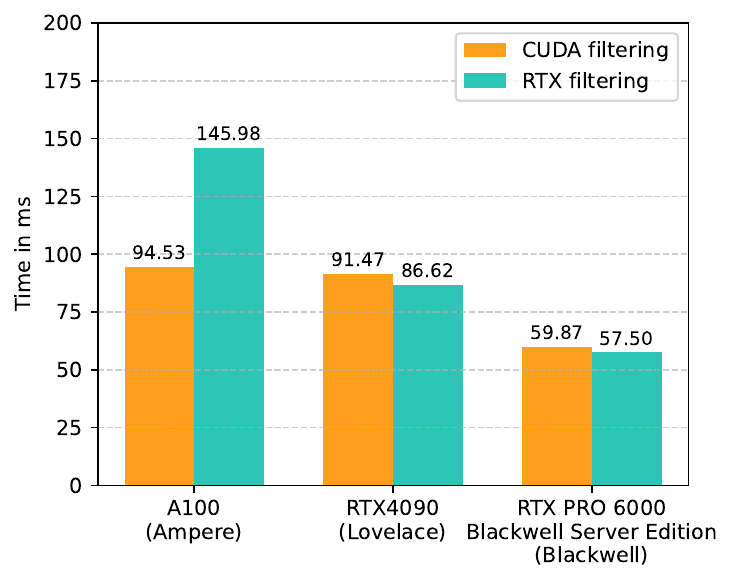}
        \caption{Testing the RT filter for different GPU architectures (Ampere, Lovelace, and Blackwell) using a uniform distribution and $2^{27}$ points.}
        \label{fig:arch}
\end{figure}

\label{subsec:future}
We also tested the algorithms on other GPUs as shown in Figure \ref{fig:arch} that shows the behavior in different architectures using a NVIDIA A100 for Ampere, NVIDIA RTX 4090 for Lovelace and NVIDIA RTX PRO 6000 Blackwell Server Edition for Blackwell, where it can be seen that there is a small advantage in the last two most recent architecture generations observed (Lovelace and Blackwell), in combination with what is shown in figure \ref{fig:arch} it can be concluded that the RT filter is highly scalable between the new generations of RT cores. It should be noted that GPUs with Ampere architecture, although these GPUs support instantiating ray tracing with the OptiX programming model for RT, do not have the hardware to directly execute the RT cores in real time, therefore they resort to traditional CUDA cores to execute the ray tracing.

\section{Conclusions}
\label{sec:conclus}
This work presents a Ray-Tracing-based algorithm to filter a set of points using RT-cores and traditional CUDA cores to accelerate the convex hull algorithm. This algorithm is designed with a three-phase algorithm where a polyhedron is first built to intersect points into a 24-face polyhedron and then points are discarded using a ƒscan-based compaction algorithm in Tensor Cores. This algorithm allows to accelerate up to $210 \times$ the filtering phase and $30 \times$ a hull algorithm with respect to a multi-core CPU implementation for a uniform distribution, and for the worst case (sphere) its performance does not degrade significantly. Moreover, it only requires $0.01\%$ of displacement of the sphere points to make the approach faster than one without filter, and $25\%$ is enough to make it faster to a multi-core CPU filtering algorithm. 

In terms of energy, both GPU filters can be considered an energy efficient approach. Both GPU filters show high energy consumption in the short term; however, the total consumption is much lower since it does not last for a long period of time like the CPU filters. On the other hand, the calculation of the convex hull is also affected by this reduction in time and energy consumption, as it significantly accelerates the calculation, having a positive impact on energy reduction.

Regarding performance scalability across different GPU architectures, it is observed that a difference persists between the use of CUDA and RT cores throughout the latest generations of GPUs.
Future work aims to implement a complete recursive version of this algorithm with dynamic BVH updates to increase the number of faces of the filtering polyhedron, in order to improve the filtering surface and find a better approximation of the convex hull using all the new features that modern GPUs can offer.

\section*{Acknowledgment}
This research was supported by the Patag\'on supercomputer~\cite{patagon} of Universidad Austral de Chile (FONDEQUIP EQM180042). This work was partially funded by 
ANID doctoral scholarship $\#21210965$, ANID FONDECYT grants $\#1241596$, $\#1221357$ and  ANID ECOS \#230017.

\printcredits

\appendix
\setcounter{table}{0}
\renewcommand{\thetable}{\thesection.\arabic{table}}
\setcounter{figure}{0}
\renewcommand{\thefigure}{\thesection.\arabic{figure}}
\section*{Appendix}

\section{Profiling Pseudohull filter algorithm}
\label{subsec:exp_cpu}
Modern CPUs  usually have two types of cores: the ones dedicated to  reach high performance and the ones that energy efficient. In this section, we want  to determine the best hardware configuration for this multi-core implementation and so to use this configuration to compared its performance against the two gpu implementations.  Table \ref{tab:exp2} shows the description of the experiments; Pseudohull 1  is to run it sequentially using 1 performance core, Pseudohull 8 is the performance using the 8 performance cores, Pseudohull 12 corresponds to run it with the 12 energy efficient cores and Pseudohull 20 corresponds to run it with all the  cores. 

\begin{table}[h!]
\begin{tabular}{llll}
    \# Total number of cores & CPU use         & Type cores \\
\hline
    \texttt{Pseudohull} $1p$   & Low          & CPU 1 core ($p$)               \\
    \texttt{Pseudohull} $8p$   & Medium       & CPU 8 cores ($p$)               \\
    \texttt{Pseudohull} $12e$  & Medium       & CPU 12 cores ($e$)              \\
    \texttt{Pseudohull} $20pe$  & High        & CPU 20 cores ($pe$)  
\end{tabular}
\caption{ Multi-core configurations to profile the first phase of the  \texttt{Pseudohull} algorithm.  The table indicates the CPU usage level and the type of core used in the CPU, where $p$ stands for performance, $e$ means energy-efficiency, and $pe$ are both type of core.}
\label{tab:exp2}
\end{table}

The \texttt{Pseudohull} algorithm consists of two phases. The first involves filtering where all points inside a tetrahedron formed by some of the endpoints are discarded, then parallel insertion of points into an array is used, and the second involves calculating the hull using a parallel implementation based on \texttt{QuickHull} within its own parallel library.

\begin{figure*}[h!] 
    \centering
    \begin{subfigure}{0.49\textwidth}
        \centering
        \includegraphics[width=1\linewidth]{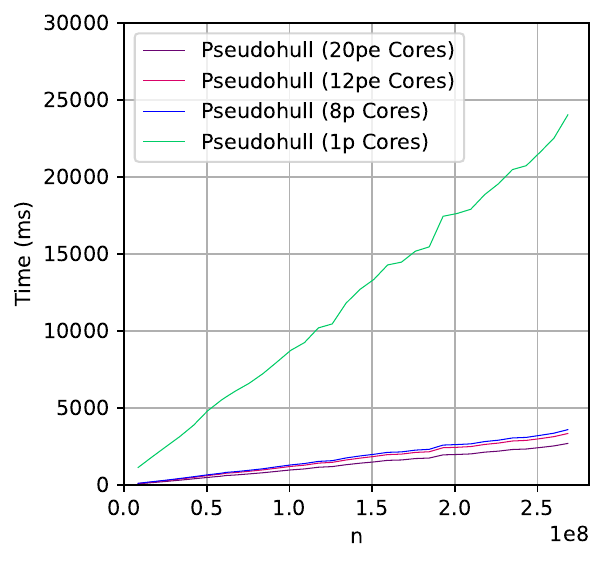}
        \caption{Uniform distribution}
    \end{subfigure}
    \hfill
    \begin{subfigure}{0.49\textwidth}
        \centering
        \includegraphics[width=1\linewidth]{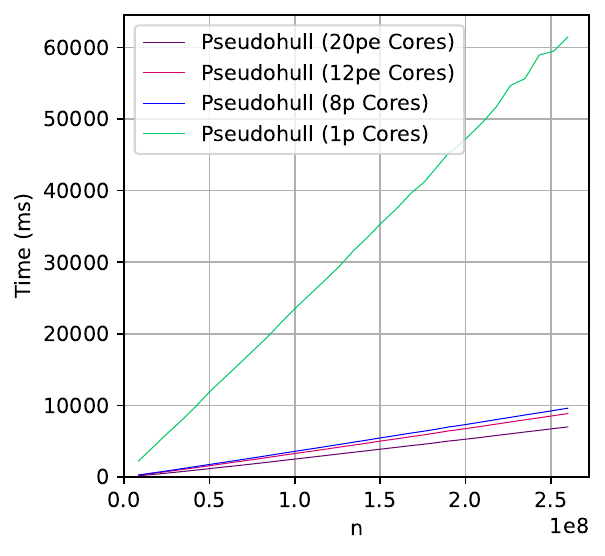}
        \caption{Sphere distribution}
    \end{subfigure}
    \caption{Plot of the filtering time for the CPU variants in both distributions.}
    \label{fig:pargeo_filter_time}
\end{figure*}

\begin{figure*}[h!] 
    \centering
    \begin{subfigure}{0.49\textwidth}
        \centering
        \includegraphics[width=1\linewidth]{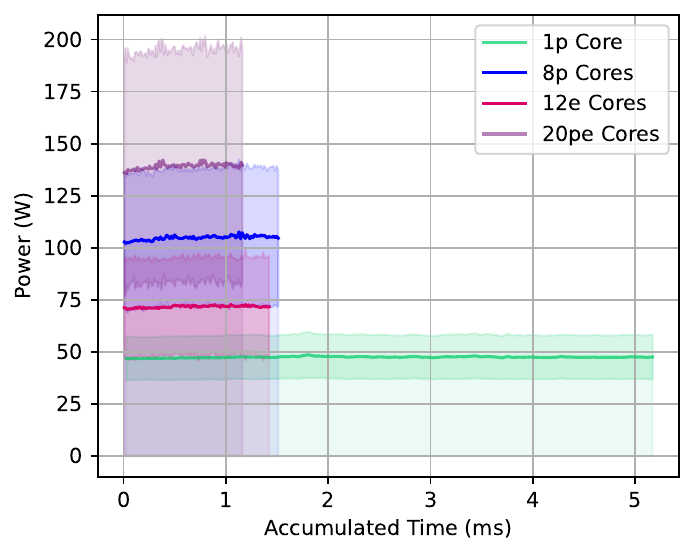}
        \caption{Uniform distribution}
    \end{subfigure}
    \hfill
    \begin{subfigure}{0.49\textwidth}
        \centering
        \includegraphics[width=1\linewidth]{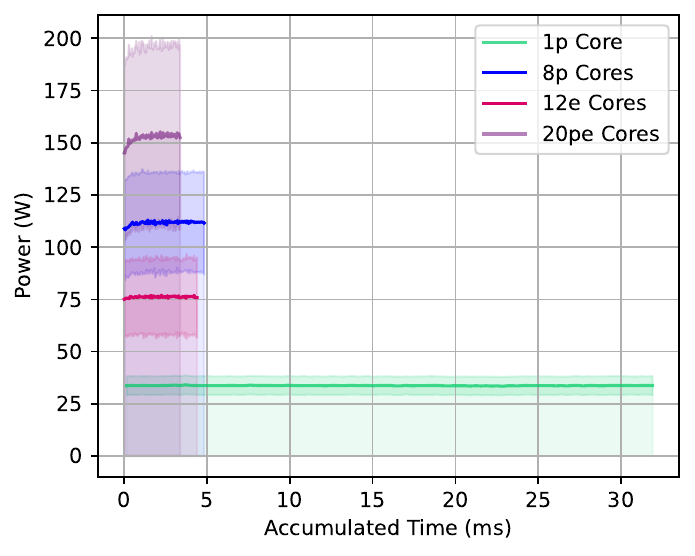}
        \caption{Sphere distribution}
    \end{subfigure}
    \caption{Energy in Watts used for all the CPU implementations over time for both distributions.}
    \label{fig:pargeo_filter_energy}
\end{figure*}

\begin{figure*}[h!] 
    \centering
    \begin{subfigure}{0.49\textwidth}
        \centering
        \includegraphics[width=0.96\linewidth]{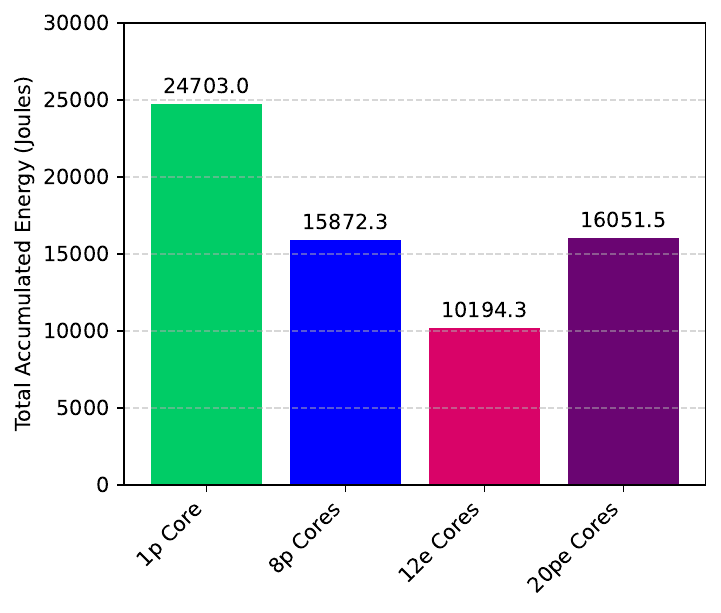}
        \caption{Uniform distribution}
    \end{subfigure}
    \hfill
    \begin{subfigure}{0.49\textwidth}
        \centering
        \includegraphics[width=1\linewidth]{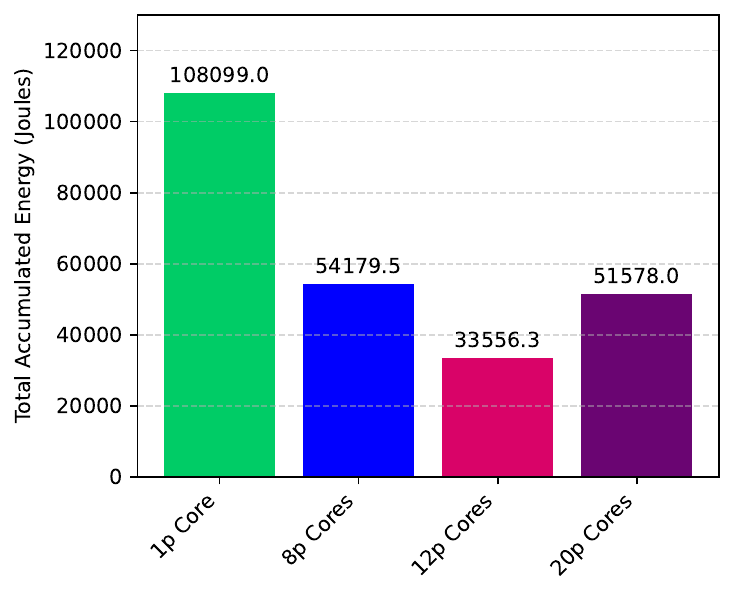}
        \caption{Sphere distribution}
    \end{subfigure}
    \caption{Accumulated energy in Joule used for all the CPU implementations for both distributions.}
    \label{fig:pargeo_filter_acc_energy}
\end{figure*}

Figure \ref{fig:pargeo_filter_time} shows the results of this benchmark for a uniform and sphere distribution, where it is possible to observe that \texttt{Pseudohull} is faster each time there are more cores available.  Counterintuitively, the multi-core configuration  using the 12 energy-efficient cores is faster than the one using the 8 performance cores, indicating that this algorithm is favored primarily by the level of parallelism rather than the clock speed of each core. However, when we observe the energy behavior in Figure \ref{fig:pargeo_filter_energy} which shows the average power used for a fixed point size where the colored area is the estimation error and the area under the curve is the energy used, in this figure it is possible to notice that the variants using efficient cores consume less power than 25\% less than the performance variants with many cores. This is also possible to note in Figure \ref{fig:pargeo_filter_acc_energy} where both parallel variants that use performance cores consume $50\sim75\%$ more total energy when computing the hull compared to the one that uses efficient cores. 

As a result of this benchmark, it is possible to conclude that for a fast filter hull computation, it is better to use all available cores, more details can be seen in Table \ref{tab:pseudohull_speedup} of Speedup with all core configurations. However, the 12 efficiency cores show an option for an energy-efficient algorithm that is fast enough for many applications that prioritize energy over speed. For the remaining experiments in this work, we utilize the 20-core variant for performance experiments and the 12-core efficiency variant for energy experiments.

\begin{table*}[h!]
    \centering
    \begin{tabular}{l|ccc|ccc}
        & \multicolumn{3}{c|}{\textbf{Uniform}} & \multicolumn{3}{c}{\textbf{Sphere}} \\
        \cline{2-7}
        \textbf{Algorithms} & $12e$ & $8p$ & $1p$ & $12e$ & $8p$ & $1p$ \\ 
        \hline
        \texttt{Pseudohull} $20pe$  & $1.24\times$ & $1.33\times$ & $9.00\times$ & $1.29\times$ & $1.40\times$ & $9.11\times$ \\
        \texttt{Pseudohull} $12e$   &      & $1.07\times$ & $7.28\times$ &      & $1.09\times$ & $7.07\times$ \\
        \texttt{Pseudohull} $8p$    &      &      & $6.78\times$ &      &      & $6.50\times$ \\
        \hline 
    \end{tabular}
    \caption{Speedup for \texttt{Pseudohull} filter phase across core configurations (Uniform vs. Sphere distributions).}
    \label{tab:pseudohull_speedup}
\end{table*}

\bibliographystyle{unsrt}

\bibliography{references}

@book{Berg:2008:CGA:1370949,
 author = {Berg, Mark de and Cheong, Otfried and Kreveld, Marc van and Overmars, Mark},
 title = {Computational Geometry: Algorithms and Applications},
 year = {2008},
 isbn = {3540779736, 9783540779735},
 edition = {3rd ed.},
 publisher = {Springer-Verlag TELOS},
 address = {Santa Clara, CA, USA},
}

@inproceedings{zhu2022rtnn,
  title={{RTNN}: accelerating neighbor search using hardware ray tracing},
  author={Zhu, Yuhao},
  booktitle={Proceedings of the 27th ACM SIGPLAN Symposium on Principles and Practice of Parallel Programming},
  pages={76--89},
  year={2022}
}

@article{zhao2023leveraging,
  title={Leveraging ray tracing cores for particle-based simulations on GPUs},
  author={Zhao, Shiwei and Lai, Zhengshou and Zhao, Jidong},
  journal={International Journal for Numerical Methods in Engineering},
  volume={124},
  number={3},
  pages={696--713},
  year={2023},
  publisher={Wiley Online Library}
}

@article{meneses2024accelerating,
  title={Accelerating range minimum queries with ray tracing cores},
  author={Meneses, Enzo and Navarro, Crist{\'o}bal A and Ferrada, H{\'e}ctor and Quezada, Felipe A},
  journal={Future Generation Computer Systems},
  volume={157},
  pages={98--111},
  year={2024},
  publisher={Elsevier}
}

@book{o1998computational,
  title={Computational geometry in C},
  author={o'Rourke, Joseph and others},
  year={1998},
  publisher={Cambridge university press}
}

@article{10.1145/321556.321564,
author = {Chand, Donald R. and Kapur, Sham S.},
title = {An Algorithm for Convex Polytopes},
year = {1970},
issue_date = {Jan. 1970},
publisher = {Association for Computing Machinery},
address = {New York, NY, USA},
volume = {17},
number = {1},
issn = {0004-5411},
url = {https://doi.org/10.1145/321556.321564},
doi = {10.1145/321556.321564},
journal = {J. ACM},
month = jan,
pages = {78–86},
numpages = {9}
}

@article{JARVIS197318,
title = {On the identification of the convex hull of a finite set of points in the plane},
journal = {Information Processing Letters},
volume = {2},
number = {1},
pages = {18-21},
year = {1973},
issn = {0020-0190},
doi = {https://doi.org/10.1016/0020-0190(73)90020-3},
url = {https://www.sciencedirect.com/science/article/pii/0020019073900203},
author = {R.A. Jarvis}
}

@article{10.1145/235815.235821,
author = {Barber, C. Bradford and Dobkin, David P. and Huhdanpaa, Hannu},
title = {The Quickhull Algorithm for Convex Hulls},
year = {1996},
issue_date = {Dec. 1996},
publisher = {Association for Computing Machinery},
address = {New York, NY, USA},
volume = {22},
number = {4},
issn = {0098-3500},
url = {https://doi.org/10.1145/235815.235821},
doi = {10.1145/235815.235821},
journal = {ACM Trans. Math. Softw.},
month = dec,
pages = {469–483},
numpages = {15}
}

@article{10.1145/359423.359430,
author = {Preparata, F. P. and Hong, S. J.},
title = {Convex Hulls of Finite Sets of Points in Two and Three Dimensions},
year = {1977},
issue_date = {Feb. 1977},
publisher = {Association for Computing Machinery},
address = {New York, NY, USA},
volume = {20},
number = {2},
issn = {0001-0782},
url = {https://doi.org/10.1145/359423.359430},
doi = {10.1145/359423.359430},
journal = {Commun. ACM},
month = feb,
pages = {87–93},
numpages = {7}
}

@article{KALLAY1984197,
title = {The complexity of incremental convex hull algorithms in Rd},
journal = {Information Processing Letters},
volume = {19},
number = {4},
pages = {197},
year = {1984},
issn = {0020-0190},
doi = {https://doi.org/10.1016/0020-0190(84)90084-X},
url = {https://www.sciencedirect.com/science/article/pii/002001908490084X},
author = {Michael Kallay}
}

@misc{cgal:hs-ch3-18b,
  author = {Susan Hert and Stefan Schirra},
  title = {{3D} Convex Hulls},
  publisher = {{CGAL Editorial Board}},
  edition = {{4.13}},
  booktitle = {{CGAL} User and Reference Manual},
  url = {https://doc.cgal.org/4.13/Manual/packages.html\#PkgConvexHull3Summary},
  year = 2018
}

@article{FERRADA2020112298,
title = {A filtering technique for fast Convex Hull construction in R2},
journal = {Journal of Computational and Applied Mathematics},
volume = {364},
pages = {112298},
year = {2020},
issn = {0377-0427},
doi = {https://doi.org/10.1016/j.cam.2019.06.014},
url = {https://www.sciencedirect.com/science/article/pii/S0377042719302870},
author = {Héctor Ferrada and Cristóbal A. Navarro and Nancy Hitschfeld}
}

@INPROCEEDINGS{5763404,  author={S. {Srungarapu} and D. P. {Reddy} and K. {Kothapalli} and P. J. {Narayanan}},  
booktitle={2011 IEEE Workshops of International Conference on Advanced Information Networking and Applications},   
title={Fast Two Dimensional Convex Hull on the GPU},
year={2011},  
volume={},  
number={},  
pages={7-12},  doi={10.1109/WAINA.2011.64}}

@article{STEIN2012265,
title = {CudaHull: Fast parallel 3D convex hull on the GPU},
journal = {Computers \& Graphics},
volume = {36},
number = {4},
pages = {265-271},
year = {2012},
note = {Applications of Geometry Processing},
issn = {0097-8493},
doi = {https://doi.org/10.1016/j.cag.2012.02.012},
url = {https://www.sciencedirect.com/science/article/pii/S0097849312000350},
author = {Ayal Stein and Eran Geva and Jihad El-Sana}
}

@article{mei,
author = {Qin, Jiayu and Mei, Gang and Cuomo, Salvatore and Sixu, Guo and Li, Yixuan},
year = {2019},
month = {04},
pages = {},
title = {CudaCHPre2D: A straightforward preprocessing approach for accelerating 2D convex hull computations on the GPU},
volume = {32},
journal = {Concurrency and Computation Practice and Experience},
doi = {10.1002/cpe.5229}
}

@inproceedings{10.1145/3350755.3400255,
author = {Blelloch, Guy E. and Gu, Yan and Shun, Julian and Sun, Yihan},
title = {Randomized Incremental Convex Hull is Highly Parallel},
year = {2020},
isbn = {9781450369350},
publisher = {Association for Computing Machinery},
address = {New York, NY, USA},
url = {https://doi.org/10.1145/3350755.3400255},
doi = {10.1145/3350755.3400255},
booktitle = {Proceedings of the 32nd ACM Symposium on Parallelism in Algorithms and Architectures},
pages = {103–115},
numpages = {13},
location = {Virtual Event, USA},
series = {SPAA '20}
}

@InProceedings{10.1007/978-3-319-94776-1_14,
author="Barbay, J{\'e}r{\'e}my
and Ochoa, Carlos",
editor="Wang, Lusheng
and Zhu, Daming",
title="Synergistic Solutions for Merging and Computing Planar Convex Hulls",
booktitle="Computing and Combinatorics",
year="2018",
publisher="Springer International Publishing",
address="Cham",
pages="156--167",
isbn="978-3-319-94776-1"
}

@article{cudach2d,
author = {Qin, Jiayu and Mei, Gang and Cuomo, Salvatore and Sixu, Guo and Li, Yixuan},
year = {2019},
month = {04},
pages = {},
title = {CudaCHPre2D: A straightforward preprocessing approach for accelerating 2D convex hull computations on the GPU},
volume = {32},
journal = {Concurrency and Computation Practice and Experience},
doi = {10.1002/cpe.5229}
}

@article{cudach3d,
author = {Mei, Gang and Xu, Nengxiong},
year = {2015},
month = {05},
pages = {35-44},
title = {CudaPre3D: An Alternative Preprocessing Algorithm for Accelerating 3D Convex Hull Computation on the GPU},
volume = {15},
journal = {Advances in Electrical and Computer Engineering},
doi = {10.4316/AECE.2015.02005}
}

@article{alshamrani,
author = {Alshamrani, Reham and Alshehri, Fatimah and Kurdi, Heba},
year = {2020},
month = {01},
pages = {317-324},
title = {A Preprocessing Technique for Fast Convex Hull Computation},
volume = {170},
journal = {Procedia Computer Science},
doi = {10.1016/j.procs.2020.03.046}
}

@inproceedings{mei2012,
author = {Mei, Gang and Tipper, John and Xu, Nengxiong},
year = {2012},
month = {12},
title = {An Algorithm for Finding Convex Hulls of Planar Point Sets},
journal = {Proceedings of 2nd International Conference on Computer Science and Network Technology, ICCSNT 2012},
booktitle = {Proceedings of 2nd International Conference on Computer Science and Network Technology, ICCSNT 2012},
doi = {10.1109/ICCSNT.2012.6526070}
}

@misc{patagon,
    howpublished = {{\url{https://patagon.uach.cl}}},
    author = {{Patag\'on Supercomputer}},
    year = {2021}
}

@inproceedings{harris2007optimizing,
  author={Harris, Mark},
  title={Optimizing {CUDA}},
  booktitle = {Proceedings of the International Conference for High Performance Computing, Networking, Storage and Analysis},
  series={SC '07},
  journal={SC07: High Performance Computing With {CUDA}},
  year={2007}
}

@inproceedings{harris_2005,
 author = {Harris, Mark},
 title = {Mapping Computational Concepts to GPUs},
 booktitle = {ACM SIGGRAPH 2005 Courses},
 series = {SIGGRAPH '05},
 year = {2005},
 location = {Los Angeles, California},
 articleno = {50},
 url = {http://doi.acm.org/10.1145/1198555.1198768},
 doi = {10.1145/1198555.1198768},
 acmid = {1198768},
 publisher = {ACM},
 address = {New York, NY, USA},
}

@inproceedings{ScanTC,
author = {Dakkak, Abdul and Li, Cheng and Xiong, Jinjun and Gelado, Isaac and Hwu, Wen-mei},
title = {Accelerating Reduction and Scan Using Tensor Core Units},
year = {2019},
isbn = {9781450360791},
publisher = {Association for Computing Machinery},
address = {New York, NY, USA},
url = {https://doi.org/10.1145/3330345.3331057},
doi = {10.1145/3330345.3331057},
booktitle = {Proceedings of the ACM International Conference on Supercomputing},
pages = {46–57},
numpages = {12},
location = {Phoenix, Arizona},
series = {ICS '19}
}

@article{NEMIRKO2021381,
title = {Machine learning algorithm based on convex hull analysis},
journal = {Procedia Computer Science},
volume = {186},
pages = {381-386},
year = {2021},
note = {14th International Symposium "Intelligent Systems},
issn = {1877-0509},
doi = {https://doi.org/10.1016/j.procs.2021.04.160},
url = {https://www.sciencedirect.com/science/article/pii/S1877050921009911},
author = {A.P. Nemirko and J.H. Dulá}
}

@article{MEERAN1997737,
title = {Optimum path planning using convex hull and local search heuristic algorithms},
journal = {Mechatronics},
volume = {7},
number = {8},
pages = {737-756},
year = {1997},
issn = {0957-4158},
doi = {https://doi.org/10.1016/S0957-4158(97)00033-0},
url = {https://www.sciencedirect.com/science/article/pii/S0957415897000330},
author = {S. Meeran and A. Share}
}

@misc{Nearchou_1994,
author="Nearchou, A. C.
and Aspragathos, N. A.",
editor="Lenar{\v{c}}i{\v{c}}, Jadran
and Ravani, Bahram",
title="A Collision-Detection Scheme Based on Convex-Hulls Concept for Generating Kinematically Feasible Robot Trajectories",
bookTitle="Advances in Robot Kinematics and Computational Geometry",
year="1994",
publisher="Springer Netherlands",
address="Dordrecht",
pages="477--484",
isbn="978-94-015-8348-0",
doi="10.1007/978-94-015-8348-0_48",
url="https://doi.org/10.1007/978-94-015-8348-0_48"
}

@misc{alanhull,
  doi = {10.48550/ARXIV.2209.12310},
  
  url = {https://arxiv.org/abs/2209.12310},
  
  author = {Keith, Alan and Ferrada, Héctor and Navarro, Cristóbal A.},
  
  title = {Accelerating the Convex Hull Computation with a Parallel GPU Algorithm},
  
  publisher = {arXiv},
  
  year = {2022},
  
  copyright = {arXiv.org perpetual, non-exclusive license}
}

@ARTICLE{9147055,
  author={Navarro, Cristóbal A. and Carrasco, Roberto and Barrientos, Ricardo J. and Riquelme, Javier A. and Vega, Raimundo},
  journal={IEEE Transactions on Parallel and Distributed Systems}, 
  title={GPU Tensor Cores for Fast Arithmetic Reductions}, 
  year={2021},
  volume={32},
  number={1},
  pages={72-84},
  doi={10.1109/TPDS.2020.3011893}}

@InProceedings{skala01,
author="Skala, Vaclav",
editor="Gervasi, Osvaldo
and Murgante, Beniamino
and Misra, Sanjay
and Garau, Chiara
and Ble{\v{c}}i{\'{c}}, Ivan
and Taniar, David
and Apduhan, Bernady O.
and Rocha, Ana Maria A.C.
and Tarantino, Eufemia
and Torre, Carmelo Maria
and Karaca, Yeliz",
title="Diameter and Convex Hull of Points Using Space Subdivision in E2 and E3",
booktitle="Computational Science and Its Applications -- ICCSA 2020",
year="2020",
publisher="Springer International Publishing",
address="Cham",
pages="286--295",
isbn="978-3-030-58799-4"
}

@article{skala02,
title = {Space subdivision to speed-up convex hull construction in E3},
journal = {Advances in Engineering Software},
volume = {91},
pages = {12-22},
year = {2016},
issn = {0965-9978},
doi = {https://doi.org/10.1016/j.advengsoft.2015.09.002},
url = {https://www.sciencedirect.com/science/article/pii/S0965997815001386},
author = {Vaclav Skala and Zuzana Majdisova and Michal Smolik}
}

@INPROCEEDINGS{skala03,
  author={Skala, Vaclav and Smolik, Michal and Majdisova, Zuzana},
  booktitle={2016 29th SIBGRAPI Conference on Graphics, Patterns and Images (SIBGRAPI)}, 
  title={Reducing the Number of Points on the Convex Hull Calculation Using the Polar Space Subdivision in E2}, 
  year={2016},
  volume={},
  number={},
  pages={40-47},
  doi={10.1109/SIBGRAPI.2016.015}}

@article{CARRASCO2024104793,
title = {An evaluation of GPU filters for accelerating the 2D convex hull},
journal = {Journal of Parallel and Distributed Computing},
volume = {184},
pages = {104793},
year = {2024},
issn = {0743-7315},
doi = {https://doi.org/10.1016/j.jpdc.2023.104793},
url = {https://www.sciencedirect.com/science/article/pii/S0743731523001636},
author = {Roberto Carrasco and Héctor Ferrada and Cristóbal A. Navarro and Nancy Hitschfeld}
}

@inproceedings{ParGeo,
author = {Wang, Yiqiu and Yu, Shangdi and Dhulipala, Laxman and Gu, Yan and Shun, Julian},
title = {ParGeo: a library for parallel computational geometry},
year = {2022},
isbn = {9781450392044},
publisher = {Association for Computing Machinery},
address = {New York, NY, USA},
url = {https://doi.org/10.1145/3503221.3508429},
doi = {10.1145/3503221.3508429},
booktitle = {Proceedings of the 27th ACM SIGPLAN Symposium on Principles and Practice of Parallel Programming},
pages = {450–452},
numpages = {3},
location = {Seoul, Republic of Korea},
series = {PPoPP '22}
}

@book{preparata1993computational,
  title={Computational Geometry: An Introduction},
  author={Preparata, F.P. and Shamos, M.},
  isbn={9780387961316},
  lccn={lc85008049},
  series={Monographs in Computer Science},
  url={https://books.google.cl/books?id=gFtvRdUY09UC},
  year={1993},
  publisher={Springer New York}
}



\end{document}